\documentclass{aa}
\usepackage{graphicx}
\usepackage{txfonts}
\usepackage{natbib}
\bibpunct{(}{)}{;}{a}{}{,}    
\newcommand{\beq}{\begin{equation}}
\newcommand{\eeq}{\end{equation}}
\newcommand{\bea}{\begin{eqnarray}}
\newcommand{\eea}{\end{eqnarray}}

\newcommand{\url}[1]{{\tt #1}}

\def\gapp{\lower 3pt\hbox{${\buildrel > \over \sim}$}\ }
\def\lapp{\lower 3pt\hbox{${\buildrel < \over \sim}$}\ }

\begin{document}
\title{Influence of an inner disc on the orbital evolution of massive planets migrating in resonance}

\author{
Aur\'elien Crida \inst{1},
Zsolt S\'andor \inst{2,3}
\and
Willy Kley \inst{1}
}
\offprints{A. Crida,\\ \email{crida@tat.physik.uni-tuebingen.de}}
\institute{
     Institut f\"ur Astronomie \& Astrophysik, 
     Universit\"at T\"ubingen,
     Auf der Morgenstelle 10, D-72076 T\"ubingen, Germany
\and
Max-Planck-Institut f\"ur Astronomie,
K\"onigstuhl 17, D-69117 Heidelberg, Germany
\and    
Department of Astronomy, E\"otv\"os Lor\'and University,
   P\'azm\'any P\'eter s\'et\'any 1/A, H-1117 Budapest, Hungary
}
\date{Received 20 December 2007 / Accepted 13 February 2008}
\abstract
{The formation of resonant pairs of planets in exoplanetary systems
involves planetary migration in the protoplanetary disc. After a
resonant capture, the subsequent migration in this configuration leads
to a large increase of planetary eccentricities if no damping
mechanism is applied. This has led to the conclusion that the
migration of resonant planetary systems cannot occur over large radial
distances and has to be terminated sufficiently rapidly through disc
dissipation.}
{In this study, we investigate whether the presence of an inner disc
might supply an eccentricity damping of the inner planet, and if this
effect could explain the observed eccentricities in some systems.}
{To investigate the influence of an inner disc, we first compute
hydrodynamic simulations of giant planets orbiting with a given
eccentricity around an inner gas disc, and measure the effect of the
latter on the planetary orbital parameters. We then perform detailed
long term calculations of the GJ~876 system. We also run N-body
simulations with artificial forces on the planets mimicking the
effects of the inner and outer discs.}
{We find that the influence of the inner disc can not be neglected,
and that it might be responsible for the observed eccentricities. In
particular, we reproduce quite well the orbital parameters of a few
systems engaged in 2:1 mean motion resonances\,: GJ~876, HD~73\,526,
HD~82\,943 and HD~128\,311. Finally, we derive analytically the effect
that the inner disc should have on the inner planet to reach a
specific orbital configuration with a given damping effect of the
outer disc on the outer planet.}
{We conclude that an inner disc, even though difficult to model
properly in hydro-dynamical simulations, should be taken into account
because of its damping effect on the eccentricity of the inner planet.
By including this effect, we can explain quite naturally the observed
orbital elements of the pairs of known resonant exoplanets.}
\keywords{Accretion, accretion discs - Planets and satellites\,: formation - Celestial mechanics}
\maketitle
\markboth
{Crida et al.\,: Inner disc and resonant planets}
{Crida et al.\,: Inner disc and resonant planets}

\section{Introduction}
\label{sec:introduction}

The orbital evolution of a system consisting of very young protoplanets 
is governed by disc-planet and mutual gravitational interactions.  In case of
differential migration the semi-major axis ratio of two planets varies
with time and - in the situation of convergent migration - capture in
a resonant configuration may occur. Indeed, a large fraction of the
observed multi-planet systems contain a pair of planets engaged in a
resonance. Here, we are interested in mean motion resonances (MMR)
where the ratio of the (mean) orbital periods of the outer to the
inner planet equals that of two small integers. Among the 6 systems
known to be in a MMR, 4 have a ratio of 2:1\,: GJ~876, HD~82\,943,
HD~128\,311 and HD~73\,526. The system discovered first to lie in a 2:1
configuration (GJ~876) is interesting in several aspects. The planets
are both very massive ($0.56$ and $1.94$ M$_{\rm Jup}$), particularly when
considering the small mass of the central star of only $0.33$
M$_\odot$. The relatively short orbital periods of the planets ($\approx 30$
and $\approx 60$ days) have allowed for very accurate determination of their
orbital elements, which are stated in Table~\ref{tab:gj876}. In GJ~876 the two
outer planets are 'deep' in the 2:1 MMR, i.e. the apsidal lines of the
two osculating orbital ellipses are always aligned and librate with
very small amplitudes only (so called apsidal resonance or corotation).  
As a consequence of the apsidal resonance the planetary eccentricities show
only small variations with time. A resonant configuration like that
in GJ~876 can only be established through the action of dissipative
effects such as disc-planet interaction. In fact, the mere existence
of systems engaged in MMRs is one of the strongest indication that
planetary migration has indeed occurred during the early evolution of
planetary systems.

The first detailed modelling of GJ~876 has been conducted by
\citet{2002ApJ...567..596L} who performed customised 3-body
simulations of a central host star and two planets with additional
(dissipative) forces mimicking the effects of disc-planet interaction.
In these type of simulations it is assumed that a pair of planets is
still embedded in the protoplanetary disc, which consists of an outer
disc only while the inner disc (inside of the inner planet) has
already been lost through effects like accretion onto star and planets
or final evaporation. In such a configuration only the outer planet
is in contact with an even further protoplanetary disc and it
experiences typically negative Lindblad torques and migrates
inward. On the contrary the inner planet has no ambient material and
does not feel any disc torque. In terms of the 3-body simulations by
\citet{2002ApJ...567..596L} this implies additional forces that reduce
the semi-major axis and eccentricity of the outer planet, while the
inner planet feels only the direct gravitational forces of the star
and outer planet. Capture into a resonant configuration can now occur
when during the inward migration the outer planet crosses the location
at which the mean orbital periods have a ratio of two small integers.

The eccentricity damping action of the outer disc onto the outer
planet is typically parameterised through the migration rate, i.e.
\beq
\frac{\dot{e}}{e} = - K \left |\frac{\dot{a}}{a} \right |
\label{eq:K}
\eeq
where $a$ and $e$ are the semi-major axis and eccentricity of the
planet, respectively, while the constant factor $K$ relates the
damping rates.  For low mass planets in the linear regime, it is known
that the eccentricity damping time scale $t_e = -e/\dot{e}$ is about
$h^2$ times shorter than the migration time scale $t_a = -a/\dot{a}$
\citep{2004ApJ...602..388T}, where $h$ is the aspect ratio of the disc
(generally a few percent). This short time scale has been confirmed
recently through fully nonlinear hydro-dynamical simulations in 2 and
3 dimensions \citep{Cresswell-etal-2007}. Hence, for low mass planets
one may safely assume that they migrate inward on nearly circular
orbits unless otherwise disturbed by additional objects in the
system. For higher planetary masses the value (and also the sign) of
$K$ is not known exactly, but due to the opening of a gap it is to be
expected that the damping on the eccentricity will be reduced with
respect to the linear case.  The simulations by
\citet{2002ApJ...567..596L} indicate now that for values of $K$ equal
to unity the resonant action leads to a strong growth of eccentricity
of both the inner and outer planets much larger than the observed
values ($e_1 \approx 0.03$ and $e_2 \approx
0.22$). \cite{2002ApJ...567..596L} pointed out that the final state of
the system is determined by $K$. For a vanishing $K$, i.e. no
eccentricity damping, the planetary eccentricities continue to
grow. To match the observations \citet{2002ApJ...567..596L} had to
assume a value of $K=100$, a value that appears to be very large
considering the high masses of the planets.

The convergent migration of two massive planets has been demonstrated
in a variety of hydro-dynamical simulations
\citep{2000MNRAS.313L..47K, 2000ApJ...540.1091B, 2001A&A...374.1092S,
2003CeMDA..87...53P}.  Through multi-dimensional hydro-dynamical
simulations of resonant planetary systems it has been shown that for
masses in the Jupiter regime the value of $K$ lies around $1-10$ at
most \citep{2004A&A...414..735K}, and in a detailed study of the
system GJ~876, \citet{2005A&A...437..727K} have shown that in
hydro-dynamical simulations where the inner disc has been depleted the
final eccentricities of the planets will always be much larger than
the observed values unless one assumes that the outer disc dissipates
rapidly on the viscous time scale. Hence, this scenario does not allow
for the migration of the resonant planets over a larger radial
distance.  While in hydro-dynamical simulations of discs with embedded
planets it is often found that the inner disc is depleted rapidly,
this may in reality not be the case and be an artefact of
inappropriate inner boundary conditions.  As shown by
\citet{Crida-Morby-2007}, even in the presence of a planet an inner
disc will survive for much longer than found previously.  In this
situation it can be expected that the inner disc will have a dynamical
influence on the inner planet and induces possibly some additional
damping of the eccentricity. The influence of such an eccentricity
damping of the inner planet was mentioned first by
\cite{2002ApJ...567..596L} for the particular case of GJ~876. They
found that a value of $K=10$ for both planets yields final
eccentricities in the observed range. The first full hydro-dynamical
study in this direction -- including an inner disc -- was done for the
resonant system HD~73\,526 by \cite{2007A&A...472..981S}\,; in their
study, they have shown that the inclusion of an inner disc indeed
leads to an additional eccentricity damping of the inner planet, and
allows more extended radial migration with reasonable final
eccentricities.

In this paper we analyse this effect in more detail and investigate
the dynamical influence that an inner disc has on a planet. In
Sect.~\ref{sec:ID} we perform a sequence of hydrodynamic simulations
and measure the torque and power exerted by the disc on the planet
and evaluate the change in eccentricity and semi-major axis as a
function of the planetary eccentricity. In Sect.~\ref{sec:gj876} we
perform a full time evolution of a pair of planets embedded in a
protoplanetary disc for realistic parameters with specific application
to GJ~876. We show that the torque and the power generated by the
inner disc yield an effective damping of $e$ which results in moderate
final eccentricities even for extended radial migration. Last, in
Sect.~\ref{sec:Nbody}, we apply an eccentricity damping to the inner
planet, in the frame of N-body simulations with artificial
non-conservative forces to mimic the effect of the disc. We apply this
to a few exoplanetary systems, and try to recover the observed
orbital elements with a realistic migration
scenario. Sect.~\ref{sec:conclusion} summaries our results.

\section{Effect of a gaseous inner disc on the orbital elements of a planet on an eccentric orbit}

\label{sec:ID}

To measure the influence that an inner disc has on the orbital
elements of a planet orbiting around it on an eccentric orbit, we
perform a suite of hydro-dynamical simulations. Here the disc is
treated as a two-dimensional gas that lies in the orbital plane of the
planet. The disc material is (basically) only present inside the
planetary orbit, so that the effect of the inner disc can be
isolated. The planet has a mass $M_p=2.14\times 10^{-3}\ M_*$ and has
a fixed orbit with semi major axis $a=1$, and a given eccentricity
$e$, that changes from one simulation to another.

The disc is treated as a non-self-gravitating gas that nevertheless
can interact gravitationally with the planet. From the expressions for
the planetary energy $E=-G(M_*+M_p)M_p/2a$ and angular momentum
$H=M_p\sqrt{G(M_*+M_p)a(1-e^2)}$, one can easily derive\,:
\begin{eqnarray}
\label{eq:adot}
\dot{a}/a & = & -\dot{E}/E\\
\dot{e}/e & = &
\frac{e^2-1}{2e^2}\left(\frac{\dot{E}}{E} + 2\frac{\dot{H}}{H}\right)
\label{eq:edot}
\end{eqnarray}
In the present simulations we keep the orbit of the planet fixed
but monitor the torque ($\dot{H}$) and power ($\dot{E}$) acting on the
planet (averaged over one orbit). We have checked that if the planet
is released from its fixed orbit, its eccentricity and semi major axis
follow the expected evolution.

The code we use is FARGO, by \citet{FARGO,FARGO2} which is a 2D
hydro-code using cylindrical coordinates ($r,\varphi$) with an
isothermal equation of state. Thus, the sound speed is given by
$c_s=hr\Omega$, where $\Omega$ is the local angular velocity, $r$ is
the distance to the central star, and $h$ is the aspect ratio, which
is here $0.05$. The gas viscosity $\nu$ is given by an
$\alpha$-prescription \citep[$\nu=\alpha c_s h r$,
][]{ShakuraSunyaev1973}, with $\alpha=10^{-2}$. The grid covers the
region from 0.4 to 1.62 in radius, divided in 112 elementary rings
(logarithmically spaced), themselves divided in 500 sectors. The inner
boundary condition is non-reflecting. More precisely, at every
time step the density in the zeroth ring is set equal to the one in
first ring, rotated by a suitable angle to mimic wave propagation,
which avoids wave reflection\,; then, the density in the zeroth ring
is shifted such that the azimuthally averaged density is the same
as initially. The outer boundary is open, which means that outflow
of gas out of the grid is permitted.

\begin{centering}
\begin{figure}
\includegraphics[angle=270,width=\linewidth]{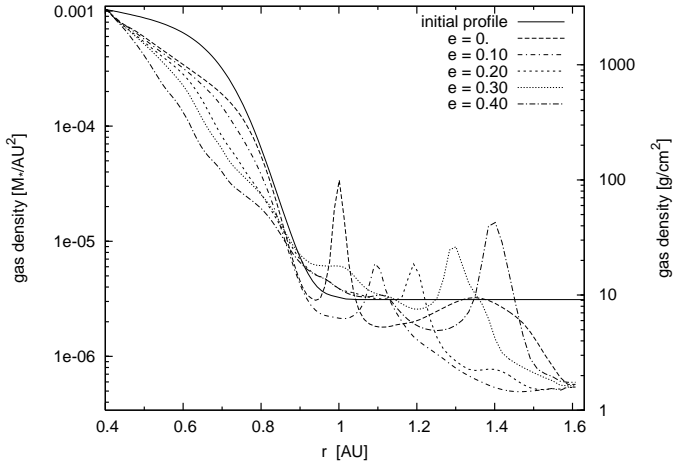}
\caption{Gas density profiles after 400 orbits for a few planet
eccentricities, and common initial profile (plain line).}
\label{fig:ID_profiles}
\end{figure}
\end{centering}

The coordinate system is centred on the star. It can be non-rotating,
corotating with the planet, or rotating at a constant angular velocity
with a period equal to that of the planet
\beq
\Omega_0 = \left(\frac{G(M_*+M_p)}{a^3}\right)^{1/2}\ .
\label{eq:Omega_0}
\eeq
Even with the FARGO algorithm, which at each ring transforms
essentially to a corotating frame, the choice of a different rotation
rate of the coordinates may change the gas dynamics close to the
planet. We have performed comparison simulations in all three frames
and display some results below. In the corotating frame, gas in the
Hill sphere simply rotates around the planet, whose motion is only
radial from apoastron to periastron and back. In the non-rotating frame,
the motion of the planet around the star in the grid tends to spread
artificially the Hill sphere. For large eccentricities, however, the
corotating frame is not advantageous because it is not rotating at a
constant angular velocity, which implies additional terms in the
equations of motion. Therefore we prefer the last option, a frame
rotating with the constant speed $\Omega_0$ in which the planet
describes an epicycle.

The initial density profile corresponds to an approximate gap opened
by the planet, without the outer disc, so that an equilibrium
profile and a stationary regime are quickly reached. The initial
profile can be seen on Fig.~\ref{fig:ID_profiles} as a solid
line. The initial total mass of gas in the disk is $1.53\times 10^{-3}
M_*=71\% M_p$. The final density profiles after 400 orbits for various
planet eccentricities are also displayed on
Fig.~\ref{fig:ID_profiles}. The profiles are taken when the planet is
at apoastron, and the density spike at $r=1+e$ corresponds to the Hill
sphere of the planet. The larger the eccentricity, the more depleted
the inner disc. The outer disc is clearly empty. In the 5 presented
cases, the simulation has been run in the non-rotating frame.

\subsection{Average effect of the inner disc}

\begin{centering}
\begin{figure}
\includegraphics[angle=270,width=\linewidth]{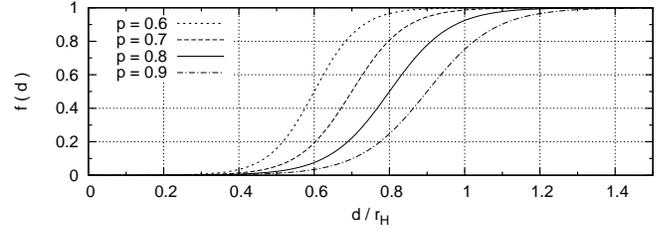}
\caption{Tapering function $f(d)$ defined by Eq.~(\ref{eq:tapering})
for four values of the parameter $p$.}
\label{fig:tapering}
\end{figure}
\end{centering}

The power and the torque from the disc on the planet become almost
constant after about 200 orbits in most cases, and we measure them
after 400 orbits. In the measure of the force exerted by the
gas on the planet, we exclude material in the Hill sphere, using a
tapering function given by\,: \beq
f(d)=\left[\exp\left(-\frac{d/r_H-p}{p/10}\right)+1\right]^{-1}
\label{eq:tapering}
\eeq
where $d$ is the distance to the planet and $r_H$ is its Hill
radius $r_H=r_p(M_p/3M_*)^{1/3}$ ($r_p$ being the distance between the
planet and the star)\,; $p$ is a dimensionless parameter set equal to
$0.8$. This function is plotted on Fig.~\ref{fig:tapering} for
different values of $p$. We stress that the use of such a tapering is
really important as the gas close to the planets exerts on it a strong
force, which should not be taken into account because this gas is
gravitationally bound to the planet and should be considered as a part
of it. With self-gravity, this gas should feel the same force as the
planet and naturally follow it\,; using a tapering frees the planet
from carrying artificially this material. The influence of the shape
of the tapering function will be discussed below.

\begin{figure}
\includegraphics[angle=270,width=0.99\linewidth]{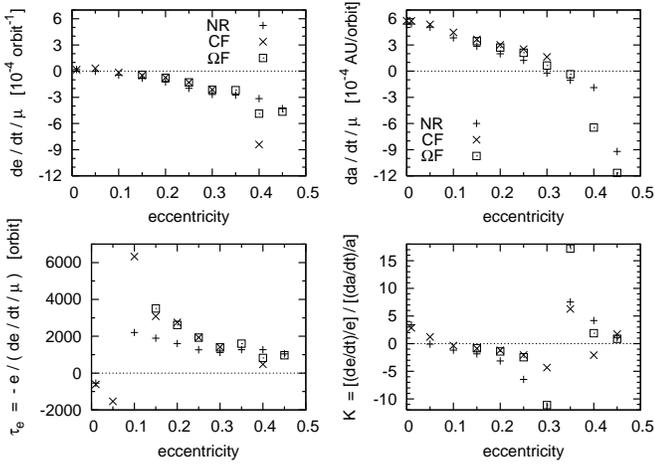}
\caption{Influence of the inner disc on a planet on a fixed orbit, as
         a function of the eccentricity. NR ($+$ symbols)\,: in the
         non-rotating frame\,; CF ($\times$ symbols)\,: in the
         corotating frame\,; $\Omega$F (open squares)\,: in the frame
         rotating monotonically with the same period as the planet.}
\label{fig:ID}
\end{figure}

The results are shown in Fig.~\ref{fig:ID} for the three above
mentioned options for the rotation rate of the coordinate system. The
quantities are normalised by $\mu=M_{\rm disc}/M_p$, because the force
felt by the planet is proportional to the gas density. The two top
panels display the influence of the inner disc on the orbital elements
$e$ and $a$. For $e\geqslant 0.1$, the eccentricity is effectively
damped on a time scale of about two thousands of orbits, as can be
seen in the bottom left panel displaying $\tau_e=-e/(\dot{e}/\mu)$\,;
the dispersion in the values of $\tau_e$ for $0.1\leqslant e \leqslant
0.15$ is simply due to the fact that $\dot{e}$ is close to zero. This
confirms the idea by \citet{Lee-Peale-2002} and
\citet{2007A&A...472..981S} that an inner disc has indeed a damping
effect on the planetary eccentricity.  For low eccentricities
($e\leqslant 0.1$), the influence of the disc is small
($|\dot{e}|<4\times 10^{-5}$ orbit$^{-1}$), and could even lead to a
small excitation of the eccentricity.

It is well-known that a planet on a circular orbit exerts a negative
torque on the inner disc \citep{LinPapaloizou1979,GT80}. It repels the
gas, leading to the opening of a gap. When the density gradient at the
gap edge is steep enough, an equilibrium arises \citep[see for
instance][]{Crida-etal-2006}. Then, the planet feels a positive torque
from the inner disc. For $e=0$, we find that the planet feels a
positive torque (which is equal to the power), as expected\,;
consequently, $\dot{a}>0$. But as $e$ increases, the power decreases,
and for $e\gtrsim 0.3$, the semi major axis variation expected if one
releases the planet is negative (top right panel). In a sense, the
inner disc attracts the planet more and more as the eccentricity
grows.

As a consequence of the variations of $\dot{a}$ and $\dot{e}$ with
$e$, the simple modelling given by Eq.~(\ref{eq:K}) with $K$ constant
seems to be an over simplification. The bottom right panel of
Fig.~\ref{fig:ID} displays $(\dot{e}/e)/(\dot{a}/a)=K$ as a function
of $e$. It varies a lot. The dispersion around $e=0.3-0.35$ is due to
the change of sign of $\dot{a}$. For $0.1<e<0.25$, a $K$ factor of the
order of minus a few seems reasonable, but a precise value can not be
stated. However, it is clear that the effect of the inner disc on the
planet can not be neglected. In particular the left panels show that a
significant damping of the eccentricity has to be taken into account.

\subsection{More detailed analysis}

\begin{centering}
\begin{figure}
\includegraphics[angle=270,width=\linewidth]{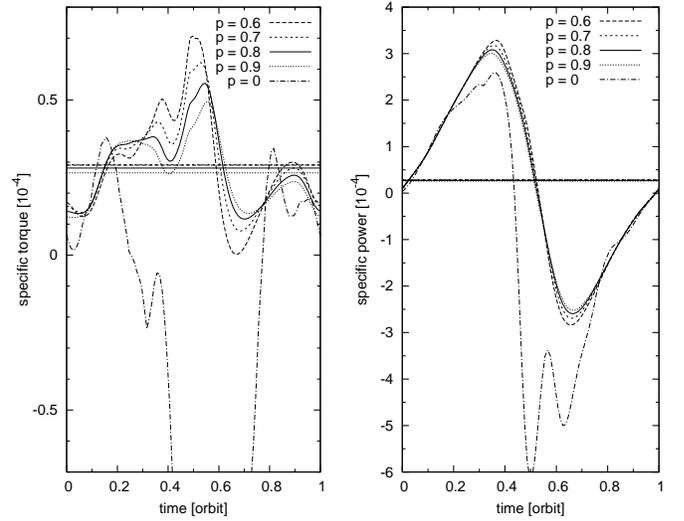}
\caption{Specific torque (left panel) and power (right panel) acting
on the planet during one orbit for a fixed semi-major axis and
eccentricity, $e=0.15$. The different curves correspond to measures
using different values of the $p$ parameter in
Eq.~(\ref{eq:tapering}). The horizontal lines correspond to the
average in each of the four cases $p\neq 0$.}
\label{fig:movie}
\end{figure}
\end{centering}

To investigate the physical mechanism more thoroughly, we focus on the
case where $e=0.15$ and where the frame is rotating at a angular
constant velocity $\Omega_0$. On Fig.~\ref{fig:movie}, we plot the
specific torque and power felt by the planet during one orbit,
starting at apoastron. The units are normalised such that $a=G=M_*=1$,
and thus $\Omega_0 = \sqrt{1.00214}$. Four curves are presented,
corresponding to measures on the same orbit with four different values
of the parameter $p$ in Eq.~(\ref{eq:tapering}), in the range
$[0.6;0.9]$. The larger the $p$ factor, the smaller the amplitude of
the oscillation. This oscillation is due to gas in the Hill
sphere. For each case, the average is plotted as a straight horizontal
line. Both the mean torque and power vary by 10 percent with varying
$p$. So does the corresponding $\dot{a}$, while $\dot{e}$ varies by
about $5\%$.

A fifth curve, labeled $p=0$, is drawn on each panel of
Fig.~\ref{fig:movie} which corresponds to the case where $f=1$ (no
tapering). Under these conditions, with an eccentric orbit and an
irregular, non-stationary density structure (also within the Hill
sphere), the measured torque and power are quite different, and
somehow noisy. The amplitude excursions are so large (into the
negative) that the mean values for the $p=0$ case, averaged over one
orbit, give $\dot{a}/\mu= -10^{-3}$ instead of $2.8\times 10^{-4}$
with $p=0.8$, and $\dot{e}/\mu= -7.14\times 10^{-4}$ instead of
$-7.95\times 10^{-5}$. This clear difference of the $p=0$ case, and
the very good agreement between the other four cases
$p\in[\,0.6\,;\,0.9\,]$, enlightens the importance and necessity of
tapering.

\begin{centering}
\begin{figure}
\includegraphics[angle=270,width=\linewidth]{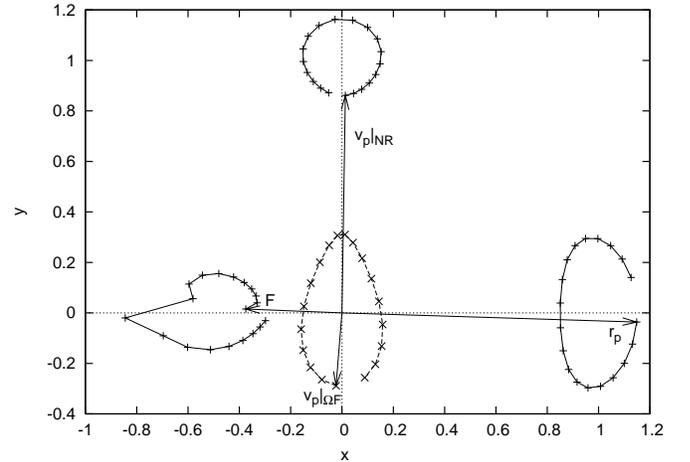}
\caption{Evolution during one orbit of the vectors $\vec{r_p}$\,:
position of the planet in the frame rotating at velocity $\Omega_0$
(right)\,; $\vec{v_p}|_{\rm NR}$\,: velocity of the planet in the
non-rotating frame (top)\,; $\vec{v_p}|_{\Omega\rm F}$\,: velocity of
the planet in the frame rotating at constant velocity $\Omega_0$
(middle)\,; $\vec{F}$\,: force from the disc on the planet (left,
arbitrary unit).}
\label{fig:orbit}
\end{figure}
\end{centering}

\begin{centering}
\begin{figure}
\includegraphics[width=\linewidth]{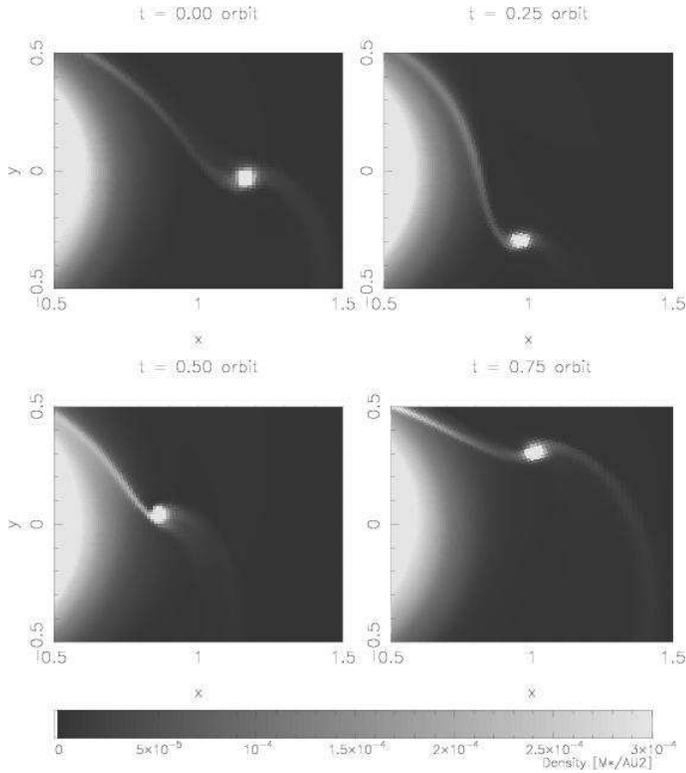}
\caption{Density maps, in linear grey scale, in the frame
rotating at constant velocity $\Omega_0$ at four different moments of
the orbit corresponding to Fig.~\ref{fig:movie}.}
\label{fig:densmulti}
\end{figure}
\end{centering}

For a better understanding of the process, on Fig.~\ref{fig:orbit} are
drawn in the frame rotating at constant angular velocity
$\Omega_0$ the traces during one orbit of the vectors $\vec{r_p}$,
$\vec{v_p}|_{\rm NR}$, $\vec{v_p}|_{\Omega\rm F}$, and $\vec{F}$
(respectively the position of the planet, its velocity in the
non-rotating and rotating frames, and the force it feels from the
disc), in the normalised units. The force is measured with $p=0.8$ and
is magnified by a factor $5\times 10^6$ for visibility. The vectors
are drawn at the beginning of the orbit\,; then, every twentieth of
orbit, a cross symbol is drawn on the curve. In the frame rotating at
the velocity $\Omega_0$, the planet describes an epicycle centred on
$(x=1,y=0)$, in the clockwise direction starting at
$(x=a+e=1.15,y=0)$. The planet velocity describes also an epicycle,
centred on $(v_x=0,v_y=a\Omega_0)$, in the anti-clockwise direction,
starting at $(v_x=0, v_y<1)$\,; the velocity in the rotating frame
$\vec{v_p}|_{\Omega\rm F}$ is plotted as a dotted line, describing a
curve around zero in the clockwise direction. As could be expected,
the force felt by the planet is directed toward the inner disc
($F_x<0$), more precisely in the direction of the wake. The wake tends
to rotate at a constant velocity. So, between apoastron and
periastron, when $y_p$ is negative (the planet is late with respect to
the average rotation at speed $\Omega_0$), the wake leads the planet
and $F_y$ is positive. After periastron, the planet leads the root of
wake, $y_p>0$ and $F_y<0$. This can be seen on the density maps of
Fig.~\ref{fig:densmulti}. This explains the variations of the sign of
the power $\vec{v_p}|_{\rm NR}.\vec{F}$ observed in
Fig.~\ref{fig:movie}. The torque $\vec{r_p}\wedge\vec{F}$ is always
positive because the angle ($\vec{r_p}$,$\vec{F}$) always remains
slightly smaller than $\pi$, as can be checked on the figure.

\subsection{Discussion}

In this section, we have treated only a particular case, with a given
planet mass and given disc parameters. The width, the depth and the
shape of the gap opened by the planet would change if the disc
viscosity, aspect ratio, or the planet mass vary, and the inner disc
effect would consequently differ. Also, some numerical issues, like
the smoothing length for the potential, the tapering, and the choice
of the frame can affect the measured force from the gas on the planet,
from a quantitative point of view. As seen from the right panel of
Fig.~\ref{fig:movie} a small change in the power, e.g. due to
numerical aspects, may possibly lead to a sign reversal of the average
value (implying a reversed migration). Exploring all the parameter
space would be prohibitive and is beyond the scope of this paper. We
simply wanted to demonstrate here that the inner disc has a non
negligible damping effect on eccentric giant planets, and we provide a
qualitative explanation of this phenomenon.


During the evolution towards the final equilibrium state the mass of
the disc is decreasing. At the end of the simulations it lies between
30 and 50 percent of the planet mass in all cases. Then the question
arises whether one could have some eccentricity excitation in the disc
induced by the planet through a tidal instability mechanism operating
through the inner 3:1 Lindblad resonance in the disc
\citep{1991ApJ...381..259L}.  In case of the inverse situation, i.e. a
planet orbiting \emph{inside} an outer disc, it has been shown that
for sufficiently large planet masses ($M_p \gapp 3 M_{\rm Jup}$) the
outer disc will turn eccentric even in case of a planet on a circular
orbit \citep{2001A&A...366..263P, 2006A&A...447..369K}.  The strength
of such a mechanism scales inversly with the planet mass such that for
the smallest planet masses the timescale of eccentricity growth is
several thousand planetary orbits.  For the instability to work the
gap created by the planet must be wide enough such that the outer 2:1
Lindblad resonance, which damps eccentricity, has been cleared.
Adding a small planetary eccentricity will reduce the necessary planet
mass slightly \citep{2006ApJ...652.1698D}.

To check for this effect in our simulations we have continued models
for $e=0.15$ for over 3500 orbits and have not seen any indication for
eccentricity growth in the disk, even in a test case where the planet
mass has been increased to $5 M_{\rm Jup}$. We can conclude that for
our planet masses we do not expect an eccentric inner disc.  Anyway,
our simulations have clearly reached a steady state when we measure
the torque and power of the force of the inner disc on the planet, and
we expect eccentricity damping of the planet by the nearly circular
disc. As pointed out above, we checked this behaviour through
simulations where we release the planet from its fixed orbit and
follow its orbital evolution self consistently. Here we find exactly
the predicted results for the change in $a$ and $e$ of the planet.

The non occurence of an eccentric inner disc can be explained by the
effect that due to the small mass of the planet and the not small
viscosity and pressure, not all the eccentricity damping resonances in
the disc are cleared, even for the large eccentricities of the planet.

\section{Application to the GJ~876 system}
\label{sec:gj876}
\subsection{Characteristics of the system}

\begin{table}
\begin{tabular}{r|cccc}
\multicolumn{5}{c}{The GJ~876 planets.}\\
\hline
\hline
name & $M.\sin i$ & period &  $a$  & $e$ \\
     & ($M_{\rm Jup}$) & (days) &  (AU) & \\
\hline
b & 1.935 & $60.93\pm 0.03$ & 0.20783 & $0.029 \pm 0.005$\\
c & 0.56 & $30.38\pm 0.03$ & 0.13 & $0.218 \pm 0.002$ \\
d & 0.023 & 1.94 & 0.020807 & 0.
\end{tabular}
\caption{Parameters of the system GJ~876 as given by
\citet{2005ApJ...622.1182L} for GJ~876\,b and GJ~876\,c, and by
\texttt{http://exoplanets.eu/} for GJ~876\,d.}
\label{tab:gj876}
\end{table}

GJ~876 is a $0.32 M_\odot$ M4 star that hosts 3 planets, the two
largest of them being in a 2:1 Mean Motion Resonance (MMR), see
Table~\ref{tab:gj876}. The resonant angles that characterise this
resonance are $\theta_1=\lambda_c-2\lambda_b+\omega_c$,
$\theta_2=\lambda_c-2\lambda_b+\omega_b$, and
$\Delta\omega=\omega_b-\omega_c$, where $\lambda$ is the mean
longitude and $\omega$ is the longitude of the periastron.  They all
three librate around $0\degr$. \citet{2005ApJ...622.1182L} estimated
the amplitude of their librations\,: $|\theta_1|_{\rm max}=7\pm 1.8
\degr$, $|\theta_2|_{\rm max}=34\pm 12\degr$, and $|\Delta\omega|_{\rm
max}=34\pm 12\degr$.

Like all the hot Jupiters, the two giant planets in GJ~876 most likely
formed further out in the protoplanetary disc (beyond the snow-line,
where water is solid and can contribute to the formation of a massive
core) and then migrated toward the central star. Migration is also
supported by the resonant motion\,: such a configuration can only be
reached by convergent motion of the planets. Thus, the outermost
planet must have been migrating inward faster than the innermost one
and eventually captured it in resonance\,; or both planets may have
been orbiting in a common gap in the disc, repelled one onto the other
by the inner and outer discs. Then, they can migrate together in this
configuration until their current position. Dissipation in the gas
disc can also modify the amplitude of libration of the resonant
angles.

The migration of two giant planets in MMR in a gas disc has been
studied first by \citet{MS2001} for the case of Jupiter and Saturn,
and in more details by \citet{Morby-Crida-2007}. They showed that if
the outermost planet is significantly lighter than the innermost one,
the migration of both planets may well be stopped or reversed outward.
In our case of GJ~876 the outer planet is substantially more massive
than the inner one, so that we expect an inward migration of the pair
of planets. Under this consideration, the observed small semi-major
axes of GJ~876\,b and GJ~876\,c are not a surprise. However, the innermost
of the two resonant planets (GJ~876\,c), being pushed inward by the outer
one (GJ~876\,b), should have its eccentricity dramatically raised,
according to N-body simulations with artificial migration rate and
detailed hydro-dynamical simulations, as discussed in the
Introduction. But we have seen that an inner gaseous disc has a
damping effect on the eccentricity of a giant planet, at least for
$e\geqslant 0.1$.  Thus, we suggest that the presence of a gaseous
disc inside the orbit of GJ~876\,c could prevent its eccentricity from
reaching unreasonable large values. To verify this hypothesis, below
we compute numerical simulations of the GJ~876 system embedded in a
disc, using a hydro-code.

\subsection{Code description and numerical parameters}
\label{subsec:prelim}

\paragraph{Numerical scheme\,:}
To have the whole disc simulated, we used the code by
\citet{Crida-etal-2007}, derived from FARGO by \citet{FARGO,FARGO2}. A
2D polar grid covers the region where the planets orbit, surrounded by
a 1D grid that extends over all the disc, the disc being assumed
axisymmetrical far from the planets. As pointed out in
\citet{Crida-etal-2007}, this 1D grid enables us to take into account
the global disc evolution, which governs the type~II migration of the
giant planets. In addition, the inner disc evolution is
self-consistently computed down to an arbitrarily small inner radius,
that could not be reached by a 2D grid for numerical reasons. Thus,
the planets can feel the influence of a realistic inner
disc. Consequently, this code is well adapted to the problem that we
wish to study.

In contrast to the previous calculations where we have used a
non-inertial frame centred on the star, the usage of the added 1D-grid
requires that the frame is inertial and centred on the centre of mass
of the system (here\,: star + planet + disc).

The tapering function used here is $f$ as given in Eq.~(\ref{eq:tapering}),
with $p=0.6$\,.

The 2D grid spans over the region where the planets orbit\,: $r\in
[0.055\,;\,0.655]$, with a resolution of $N_s = 500$ sectors in
azimuth and $N_r=300$ rings in radius\,; the rings width is thus
$\delta r = 0.002$~AU, which also applies for the 1D grid. The outer
edge of the 1D grid is arbitrarily fixed at $r=10$~AU, which is far
enough. The inner edge will be discussed below.

\paragraph{Disc parameters\,:}
The damping effect of the inner disc is proportional to its mass, in
particular to the gas density and the amplitude of the wake close to
the planet. The deeper the gap opened by the innermost planet, the
smaller the wake there. The wider the gap, the further the disc lies
from the planetary orbit. Consequently, a larger gap leads to a
smaller damping. The shape of the gap is determined by the gas
parameters $\nu$ and $h$ \citep{Crida-etal-2006}. Thus, these
parameters play a crucial role for the damping as well. Here, the gas
viscosity is still given by an $\alpha$-prescription, with
$\alpha=10^{-2}$. The chosen aspect ratio is $h=0.07$.

\paragraph{Radius of the inner edge of the disc\,:}
\citet{Crida-Morby-2007} have shown that the inner disc evolution is
strongly dependent on the radius of the inner edge of the disc, and
more precisely on the ratio between this radius and the radius of the
planetary orbit. Generally, this radius is poorly constrained, and
strongly model-dependent. So, the mass of the inner disc could be very
uncertain\,: shall this radius be close to $0.13$~AU, and there would be
no inner disc in GJ~876 ; shall it be close to the stellar radius and
there would be room for a massive inner disc.

Fortunately, in the case of GJ~876, a `hot Neptune' is present at
$0.02$~AU. This gives a strong constraint for the location of the
inner boundary. Indeed, the migration of this planet stopped there for
some reason.

A first possibility is that this planet, not massive enough to open a
gap, and thus migrating in the type~I regime, was caught in a planet
trap \citep{Masset-etal-2006}. This planet trap could be the inner
edge of the disc\,: indeed at this location, the disc density
increases rapidly from zero, leading most probably to a positive
gradient of the vortensity and a strong corotation torque. Then, the
radius of the inner edge of the disc should have been close to $0.02$~AU.

A second possibility is that GJ~876\,d migrated inwards in the gas disc
until it lies in the empty cavity inside the edge of the disc. A few
reasons may explain that this planet crossed the planet trap created
by the disc edge. The trap may not have been strong enough (a jump in
density over a only few scale heights of the disc is required) ; also,
the aspect ratio and the viscosity of the disc may have been so low at
this place that this planet could perturb the disc profile destroying
the trap. Finally, turbulence in the disc causes random torques,
helpful to jump the trap. Once the planet is in the cavity, it feels a
negative torque from the disc at outer Lindblad resonances
\citep{GT79,GT80}. Thus, the planet goes on migrating inward until
there is no more gas at the location of its 2:1 resonance (the
outermost one). In that case, the inner edge of the disc must have
been located at the 2:1 resonance with GJ~876\,d, that is at
$0.033$~AU. Then, as the disc is evaporated by the star from inside
out, the planet remains there.

Let us focus on this second possibility. We claim that the inner disc
could not extend further inward than $0.033$~AU, otherwise it would
have pushed GJ~876\,d inward, and that it should have extended exactly
down to this radius, otherwise there would be no reason for GJ~876\,d to
migrate inward to its present position. Thus, the open inner edge of
the 1D grid will be located at $r=0.033$ in our simulation.

\subsection{Results}
\begin{figure}
\includegraphics[angle=270,width=0.99\linewidth]{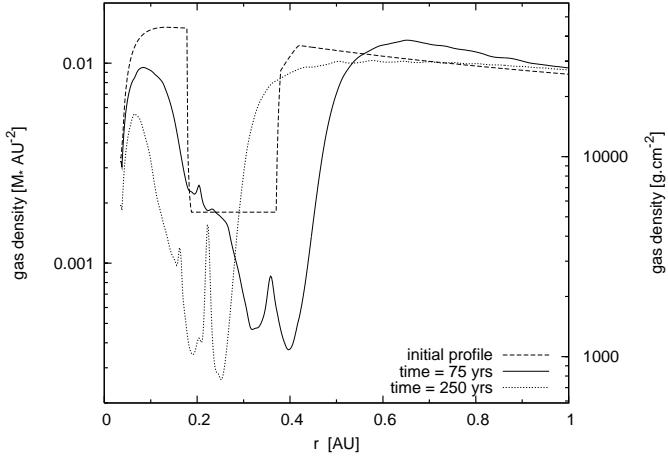}
\caption{Density profiles at time 0 (dashed line), at the moment where
the planets are released (solid line), and at the end of the
simulation (after 250 years, dotted line).}
\label{fig:profiles2}
\end{figure}

The innermost planet, GJ~876\,d, is not computed. The two largest ones
are launched on circular orbits at $r=0.36$ and $0.21$~AU
respectively. At first, the planets influence each other and influence
the disc, but do not feel the disc effect. So, the planets shape a gap
in the density distribution and the gas disc reaches an equilibrium
for this planetary configuration. This phase lasts for 75 years, which
makes about 200 orbits for the outer planet. The initial density
profile as well as the profile at this time are shown on
Fig.~\ref{fig:profiles2}\,; the final density profile is also
plotted. The mass of gas present in the 2D grid after this first
phase is $2.69\times 10^{-2}$ stellar mass ($\sim 1.7\times 10^{28}$
kg).

\begin{figure}
\includegraphics[angle=270,width=0.99\linewidth]{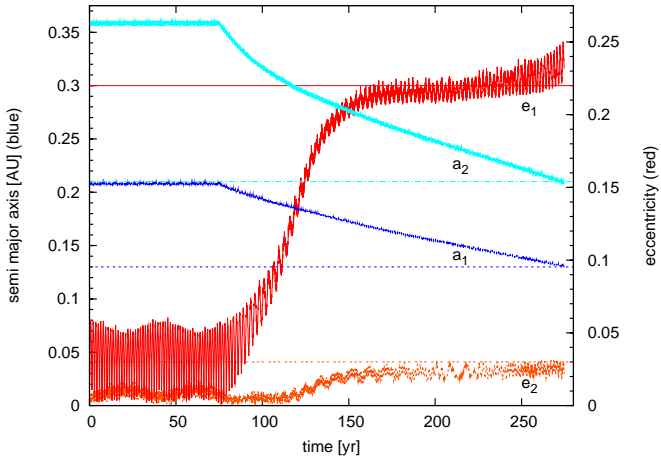}
\caption{Semi major axis (blue, left $y$-axis) and eccentricity (red,
right $y$-axis) evolution of GJ~876\,b (light colour, $a_1$, $e_1$) and
GJ~876\,c (dark, $a_2$, $e_2$). The horizontal lines correspond to
the observed values, as given by Table~\ref{tab:gj876}.}
\label{fig:GJ876_Aurel}
\end{figure}

Then, the planets are released and allowed to move under the influence
of the gas. The evolution of their orbital elements is shown on
Fig.~\ref{fig:GJ876_Aurel}. As expected, the outer planet migrates
inward, pushed by the outer disc (curve labelled $a_2$ on
Fig.~\ref{fig:GJ876_Aurel}). However, the inner planet also migrates
inward, although less rapidly. This is because it did not open a very
clean gap (see Fig.~\ref{fig:profiles2}) and it lies on the inner edge
of the gap opened by GJ~876\,b\,; consequently the inner planet feels a
strong negative corotation torque.

\begin{figure}
\includegraphics[angle=270,width=0.99\linewidth]{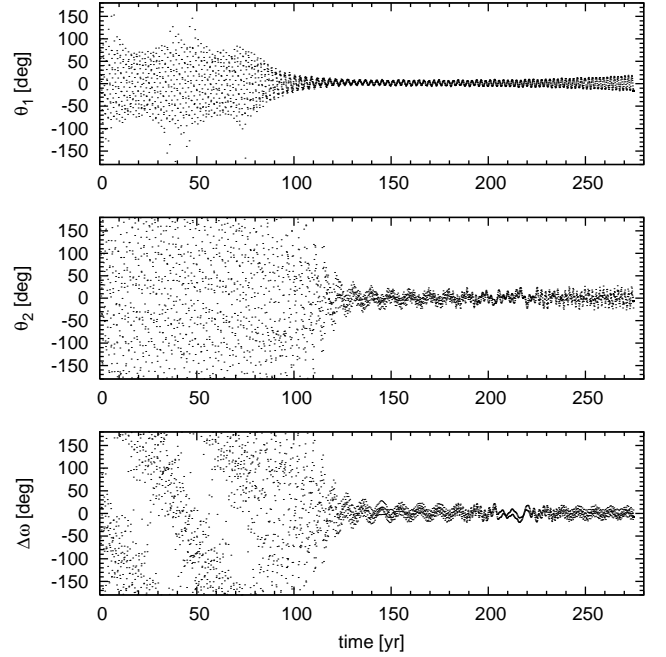}
\caption{Resonant angles associated with the 2:1 resonance
$\theta_1$, $\theta_2$, $\Delta\omega$ as functions of time.}
\label{fig:resonance}
\end{figure}

The 3 relevant resonant angles associated with the 2:1 resonance are
shown on Fig.~\ref{fig:resonance}. After $\sim 110$ years, the outer
planet catches the inner one in its 1:2 Mean Motion Resonance
($\theta_1$, $\theta_2$ and $\Delta\omega$ start librating around
$0\degr$ with small amplitude), and the pair of planets goes on
migrating in this configuration. The eccentricities rise, as
expected. But after a phase of eccentricity growing, a limit value is
reached for $e_1$ and $e_2$ at about 150 years. The planets go on
migrating at the same rate, while their eccentricities remain
constant. The value obtained for the eccentricities is very close to
the one given by \citet{2005ApJ...622.1182L}. After $\sim 270$ years,
the planets have reached their present semi major axes, and their
eccentricities are oscillating in the ranges $0.019<e_2<0.032$ and
$0.21<e_1<0.25$. The amplitude of the libration of the resonant angles
at this point are $|\theta_1|_{\rm max}\approx 18\degr$ (but it was
about $8\degr$ at 220 years), $|\theta_2|_{\rm max}\approx 28\degr$,
and $|\Delta\omega|_{\rm max}\approx 20\degr$. These amplitudes are
somewhat larger than the ones given by \citet{2005ApJ...622.1182L},
but the agreement for the semi majors axes and eccentricities is
excellent.

\subsection{Discussion}

\begin{figure}
\includegraphics[angle=270,width=0.99\linewidth]{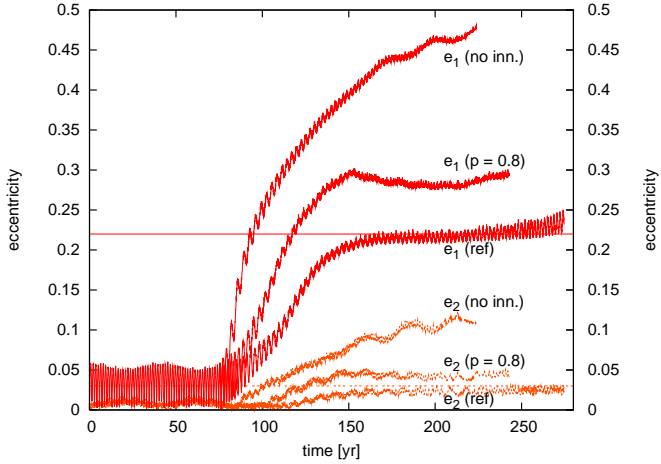}
\caption{Eccentricity evolution of GJ~876\,b (light colour, $e_2$) and
GJ~876\,c (dark, $e_1$) in three cases\,: (\emph{i}) standard
simulation (same as Fig.~\ref{fig:GJ876_Aurel}, curves labelled
``ref'')\,; (\emph{ii}) same as ({\it i}) but the inner planet is not
affected by the disc (curves labelled ``no inn.'')\,; ({\it iii}) same
as (\emph{i}) but the $p$ parameter of the tapering function $f$ is
$0.8$ (curves labelled ``p = 0.8''). The horizontal lines correspond
to the observed values, as given by Table~\ref{tab:gj876}.}
\label{fig:GJ876_InnerNO}
\end{figure}

\paragraph{Role of the inner disc\,:}
For comparison, the same simulation has been run with no action of the
disc on the inner planet. The semi major axis evolution of the two
planets is almost not affected\,; indeed migration is dominated by the
outermost planet, which is pushed inward by the outer disc, and that
pushes inward the inner planet through the resonance
locking. Consequently, the curves of $a_i(t)$ overlap the ones of
Fig.~\ref{fig:GJ876_Aurel}. On the contrary, the behaviour of the
eccentricity of the inner planet $e_1$ changes dramatically. It
increases faster and continuously, as expected from N-body
simulations. In fact, both eccentricities rise up to high values that
are not compatible with the observations ($e_1\sim 0.45$,
$e_2>0.1$). The eccentricities evolution is
displayed on Fig.~\ref{fig:GJ876_InnerNO} for the previous case
(labelled ``ref''), and in the case where the action of the disc
on the inner planet is switched off (labelled ``no inn.''). This
convincingly shows the very important role of the inner disc in
eccentricity damping for the inner planet, which in turn affects the
eccentricity of the outer one as well.

\paragraph{Role of the tapering function\,:}
We should also mention that in the standard simulation, if one takes
$p=0.8$ instead of $0.6$ in the tapering function $f$ (see
Eq.~(\ref{eq:tapering}) and Fig.~\ref{fig:tapering}), the inner planet
reaches a higher limit eccentricity (between 0.275 and 0.3), while the
eccentricity of the outer one saturates between 0.035 and 0.05. The
eccentricities evolution in this case is also displayed on
Fig.~\ref{fig:GJ876_InnerNO} (curves labelled ``p = 0.8''). The
eccentricities still do not reach extremely high values thanks to the
disc damping, but this is not in very good agreement with the
observations (horizontal lines on the figure), especially for what
concerns the inner planet. This also shows that the tapering function
can have an influence on the final eccentricity of the planets in
numerical simulations (as could be expected from Fig.~\ref{fig:movie}),
and one should take this into consideration.

\paragraph{Disc dispersal\,:}
To match with the present state, one has to disperse the disc when the
planets have reached their present semi-major axis (at time $t \sim
270$ yr). This is a common problem when modelling the extra-solar
planets.

\begin{figure}
\includegraphics[angle=270,width=0.99\linewidth]{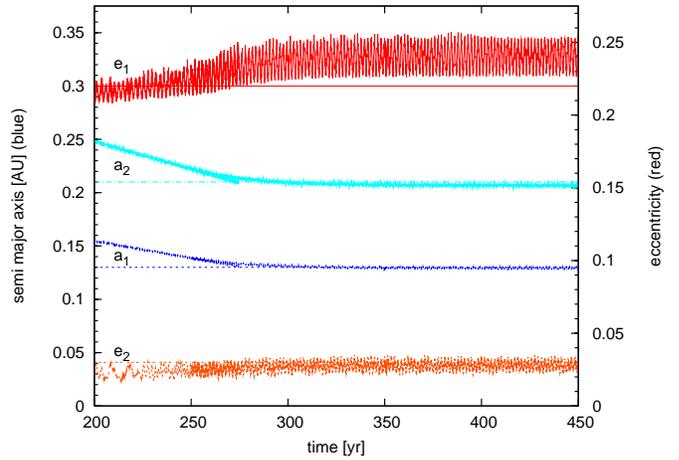}
\caption{Semi major axis and eccentricity evolution of GJ~876\,b and
GJ~876\,c, with disc clearing from time $t=250 yr$ on. The horizontal
lines correspond to the observed values, as given by
Table~\ref{tab:gj876}.}
\label{fig:November9_D_damp}
\end{figure}

To check what happens when one removes the gas disc, we applied a
procedure also used in \citet{Morby-etal-2007} and
\citet{Thommes-etal-2007} to have the gas disappear smoothly for the
planetary system not to be perturbed by a sudden change in the
potential\,: from time $t=250 yr$, the gas density is decreased
exponentially in each cell of the grid with a time scale of $27.5$
years. The result is shown on Fig.~\ref{fig:November9_D_damp}\,: as
the gas disappears, the planets remain in 2:1 MMR, while their
migration speed decreases exponentially (it is well known that when
the planet is more massive than the disc, the inertia of the planet is
the limiting factor in the type~II migration regime). One should note
in passing that as the migration speed and the eccentricity damping
are both proportional to the gas density, the $K$ factor is not
affected in this procedure\,; consequently, the equilibrium value of
the eccentricities is not affected, and they remain close to 0.03 and
0.22 respectively, while the semi major axes converge to 0.21 and
0.13~AU. In the end, gas has almost completely disappeared, and we are
left with a planetary system very similar to GJ~876.

The disc clearing process is complex and not well constrained, in
particular in the vicinity of the star. In general, the gas density
first slowly decreases while the disc accretes onto the central star
and spreads outward. When the density is low, the photo-evaporation by
the central star can play a significant role and erode the disc. The
extreme UV photons ionise and heat the upper layer of the disc, so
that gas can leave the potential well of the star, and the density
decreases. However, this works more efficiently at radius larger than
about $1$~AU (close to the star, the gravity is too
strong). Consequently, the region where the two giant planets of GJ~876
are orbiting should not be much affected\,; in particular, the inner
disc should not disappear. The remnant disc inside $\sim 1$~AU will
viscously spread onto the star and out. So, the migration path of the
planets should not be much affected. In addition, the viscous
evolution dominates the disc evolution until the density is very
low. Then, photo-evaporation happens on a timescale that is much
shorter than the disc life time. Thus, the final phase of gas
dispersal occurs when the disc is too low to have any significant
influence on the planets on such a short timescale.

Thus, we think that the planetary configuration should not be
significantly perturbed during this phase, and that the above
modelling by exponential damping of the density gives reliable
results. Anyway, these planets somehow migrated toward their
host star, until the gas density was too low to push them. They were
left at 0.13 and 0.21~AU, and then the disc disappeared. Our
simulation shows that when the planets stop migrating, they
automatically have the correct eccentricities\,; hence, our simulation
is consistent with the observed configuration. To our knowledge, this
is the first time that an extra-solar system is reproduced in a fully
hydro simulation taking into account all the protoplanetary disc
and allowing for a significant migration of the planets with correct
eccentricities.

\section{Modelling an inner disc by N-body calculations}
\label{sec:Nbody}

Studying the effect of an inner disc on the evolution of the resonant
planetary systems by full hydro-dynamical calculations (as done in the
previous section) requires typically a large amount of computer
time. Fortunately, the effects of the outer and inner discs can also
be modelled approximately conveniently by gravitational N-body
simulations using properly parameterised non-conservative
forces. These forces can be derived by using the migration rate $\dot
a/a$ and the eccentricity damping rate $\dot e/e$ of a planet embedded
into the protoplanetary disc \citep[see][ for two different
approaches]{2002ApJ...567..596L,2006MNRAS.365.1160B}. Instead of the
migration and damping rates one can also use the corresponding
$e-$folding times defined as $\tau_a = -(\dot a/a)^{-1}$ and
$\tau_e=-(\dot e/e)^{-1}$.

When studying the formation of a resonant system consisting of an
inner and an outer giant planet, usually the outer planet is forced to
migrate inward. When the ratio of their semi-major axes approaches a
critical limit, a resonant capture may take place between them
\citep[for the conditions of a resonant capture into the 2:1 MMR
see][]{kleysand07}. After the resonant capture the two planets migrate
inward as the outer one still feels the negative tidal torques of the
disc.

Having studied in the previous section the hydro-dynamical evolution
of GJ~876 thoroughly, in this section we provide further results
obtained in the framework of the three-body problem with dissipative
forces. Our results show that the presence of an inner disc during the
migration of the giant planets is consistent with the observed state
of the resonant systems having two giant planets engaged in a 2:1 MMR.

Since the eccentricity of the inner planet is excited by the resonant
interaction, its orbit becomes more and more elongated penetrating
into the inner disc. This represents a damping mechanism which acts
against the eccentricity excitation, and may set a quasi-equilibrium
between these processes keeping the eccentricity of the inner planet
on a constant value (within certain limits) during the whole migration
process. The effect of an inner disc can be investigated by using a
repelling non-conservative force acting on the inner planet
parameterised by a positive migration rate $\dot a/a$ and a negative
$\dot e/e$. Similarly to the case of the inward migrating outer
planet, a ratio $K$ of the above parameters can also be
defined. According to the definition of the $e-$folding times,
$\tau_a$ will have a negative sign in this case.

We note, however, that when both the outward and inward migration of
the outer and inner planets are considered simultaneously, the final
state of the system cannot be characterised uniquely by using only the
$K$ ratios for the inward and outward migration.  The final values of
the semi-major axes and the eccentricities depend directly on the
migration rates (or on the corresponding $e-$folding times) $\dot
a_1/a_1$, $\dot a_2/a_2$ and the eccentricity damping rates $\dot
e_1/e_1$, $\dot e_2/e_2$ (where indices '1' and '2' stay for the inner
and outer planet, respectively), and not only on their ratios $K_1$
and $K_2$.  The characterisation of the system's final state by using
the migration parameters will be discussed more detailed at the end of
this section.

In the following we repeat the three-body calculations for GJ~876 by
\cite{2002ApJ...567..596L} adding the effects of an inner disc, then
we review the simulations for HD~73\,526 by \cite{2007A&A...472..981S},
and finally present our new results in modelling the formation of
HD~82\,943 and HD~128\,311 with an inner disc.

\subsection{GJ~876 and HD~73\,526}

\begin{figure}  
   \centering  
   \includegraphics[width=8cm]{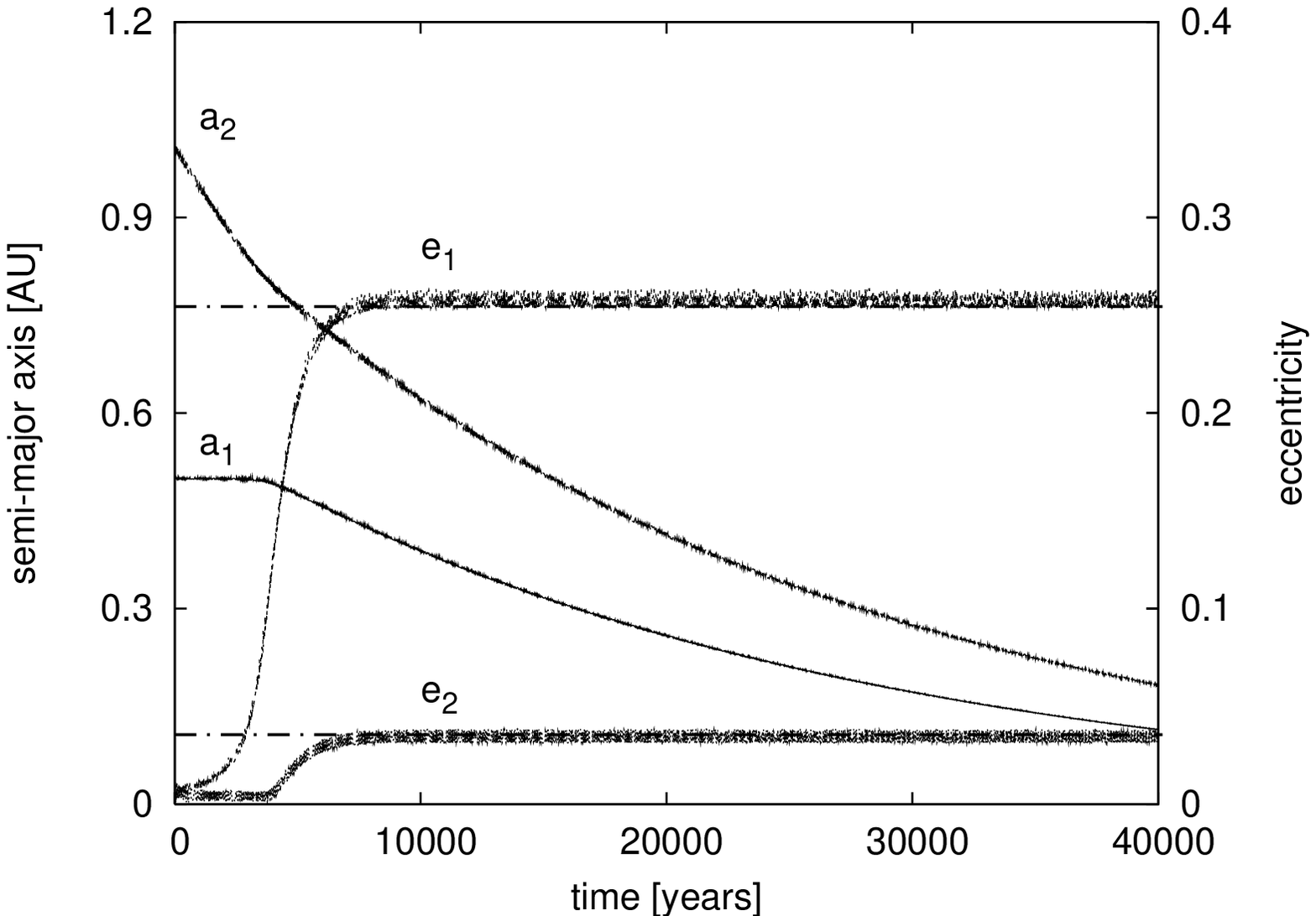}  
   \centering  
   \includegraphics[width=8cm]{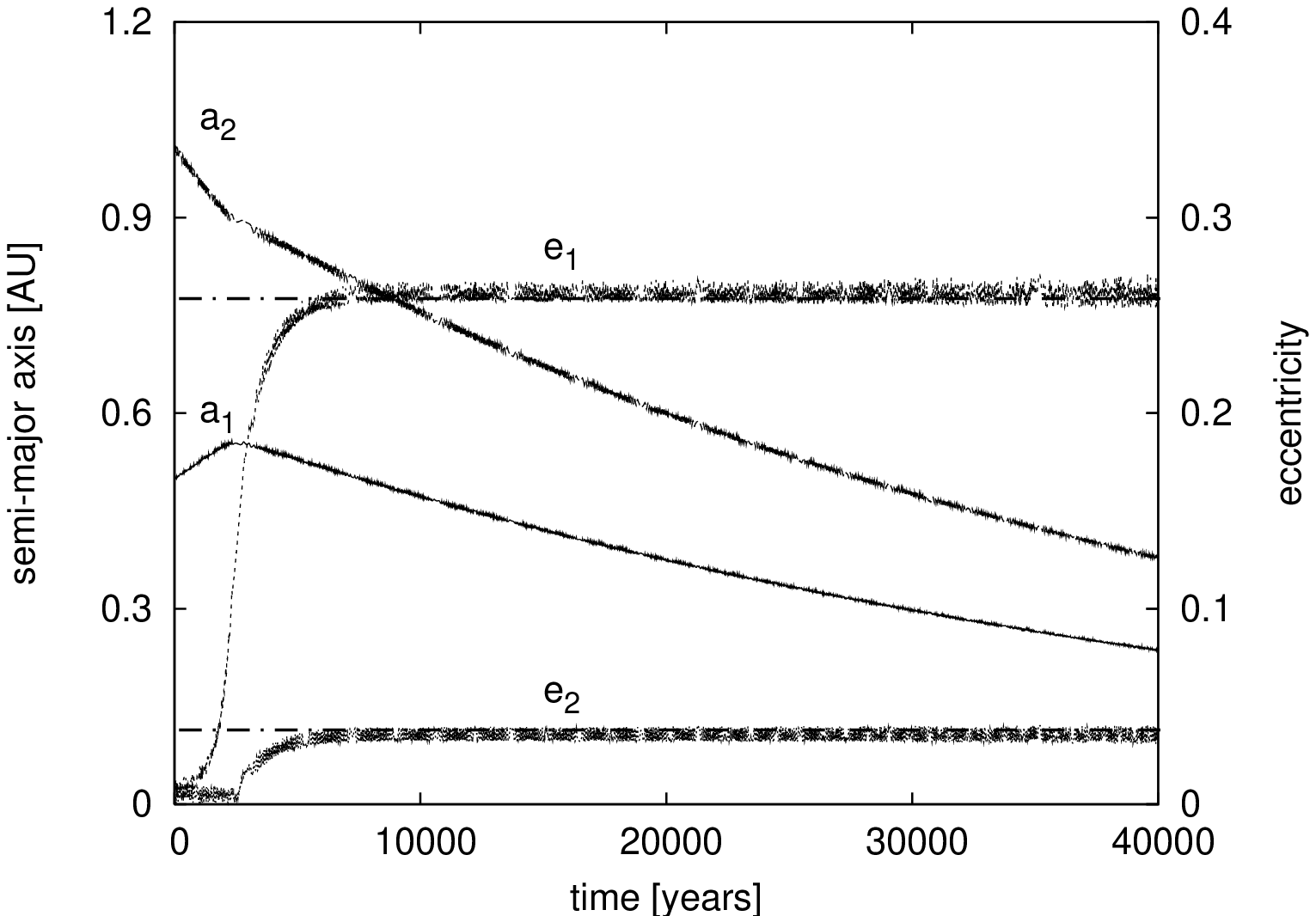}  
      \caption{Behaviour of the semi-major axes and the eccentricities
      of the resonant giant planets in the system GJ~876. The
      horizontal lines correspond to the observed values of the
      eccentricities. \emph{Top\,:} Only the outer disc is taken into
      account, and therefore only the orbital evolution of the outer
      planet is affected, with the $e-$folding times
      $\tau_{a_2}=2\times 10^4$ years, $\tau_{e_2}=2\times 10^2$ years,
      so $K_2=100$. \emph{Bottom\,:} Same as above in the presence of
      outer and inner discs. In this case, $\tau_{a_1}=-2\times 10^4$,
      $\tau_{e_1}=2.5\times 10^3$, $\tau_{a_2} = 2\times 10^4$,
      and $\tau_{e_2}=2.5\times 10^3$ years (giving $K_1=K_2=8$).}
      \label{fig:gj876_diss_nbody}  
\end{figure}

\paragraph{GJ~876\,:}
First we repeated the simulations of \cite{2002ApJ...567..596L} on the
formation of GJ~876 by using three-body calculations with dissipative
forces. Similarly to their results, we required
$K_2=\tau_{a_2}/\tau_{e_2}=100$ if only the outer giant was affected by an
outer disc.  The evolution of the semi-major axes and the
eccentricities is shown in the top panel of
Fig.~\ref{fig:gj876_diss_nbody}. In our calculations we used
$\tau_{a_2}=2\times 10^4$ and $\tau_{e_2}=2\times 10^2$ years giving
exactly $K_2=100$. We note again that this ratio between the
$e-$folding times is very high and maybe physically unrealistic for
massive planets.

To quantify the damping effect of the inner disc on the inner planet
while the giant planets revolve in a common gap, we performed
further three-body simulations with also $\tau_{a_1}=-2\times 10^4$
and $\tau_{e_1}=2.5\times 10^3$ years, where we point out that the
minus sign of $\tau_{a_1}$ stands for the outward migration. To model
the damping effect of the outer disc on the outer planet we used the
$e-$folding times $\tau_{a_2} = 2\times 10^4$ and
$\tau_{e_2}=2.5\times 10^3$ years. Note that the above migration
parameters correspond to $K_1=K_2=8$. The result of our calculations
is shown in the bottom of Fig.~\ref{fig:gj876_diss_nbody}. These
figures in fact are very similar to those obtained by
\cite{2002ApJ...567..596L}, however to model the behaviour of the
system, we used different migration parameters.

\paragraph{HD~73\,526\,:}
As we already mentioned, the effect of an inner disc was studied more
detailed by \cite{2007A&A...472..981S} in the case of
HD~73\,526. There, it has been shown that if the outer planet is
damped only, one needs $K_2 = 15-20$, which according to the
hydro-dynamical simulations seems to be a too high value. Moreover, if
only the outer planet is damped, the eccentricity of the inner planet
shows an increasing tendency, which can result in exceeding the limit
coming from the observations ($e_{\rm inn} \approx 0.3$). On the other
hand, if in addition the inner planet is damped by an inner disc, the
eccentricities will level off and the outcome of the migration
scenario fits better to the observations\,: the formation of
HD~73\,526 could be modelled successfully by using the $e-$folding
times $\tau_{a_1}=-5\times 10^4$, $\tau_{a_2}=10^4$ years and
$\tau_{e_1}=5\times 10^3$, $\tau_{e_2}=10^3$ (corresponding to
$K_1=K_2=10$). The behaviour of the eccentricities in this case
is shown in the top panel of Fig.~8 in the paper of
\cite{2007A&A...472..981S}.

We can conclude that the presence and effect of an inner disc in
modelling the formation processes of GJ~876 and HD~73\,526 is physically
more realistic, giving results fitting very well to the radial velocity
observations. In what follows, we show that in the case of the systems
HD~82\,943 and HD~128\,311 the presence of an inner disc is still a real
alternative, however these systems do not need such strong dampings
for the eccentricity of their inner giant planets.

\subsection{HD~82\,943 and HD~128\,311}

\begin{figure}  
   \centering  
   \includegraphics[width=8cm]{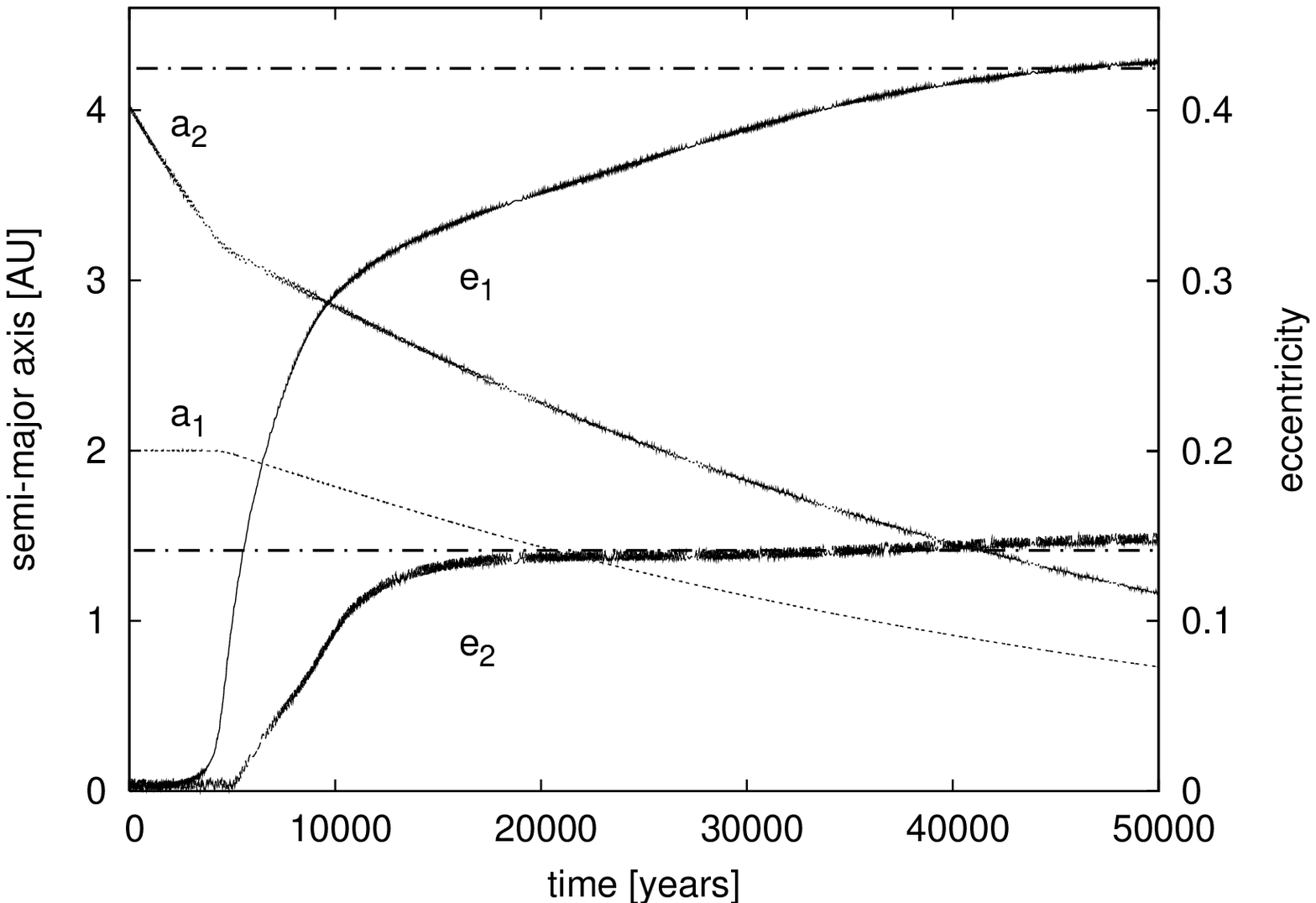}  
   \centering  
   \includegraphics[width=8cm]{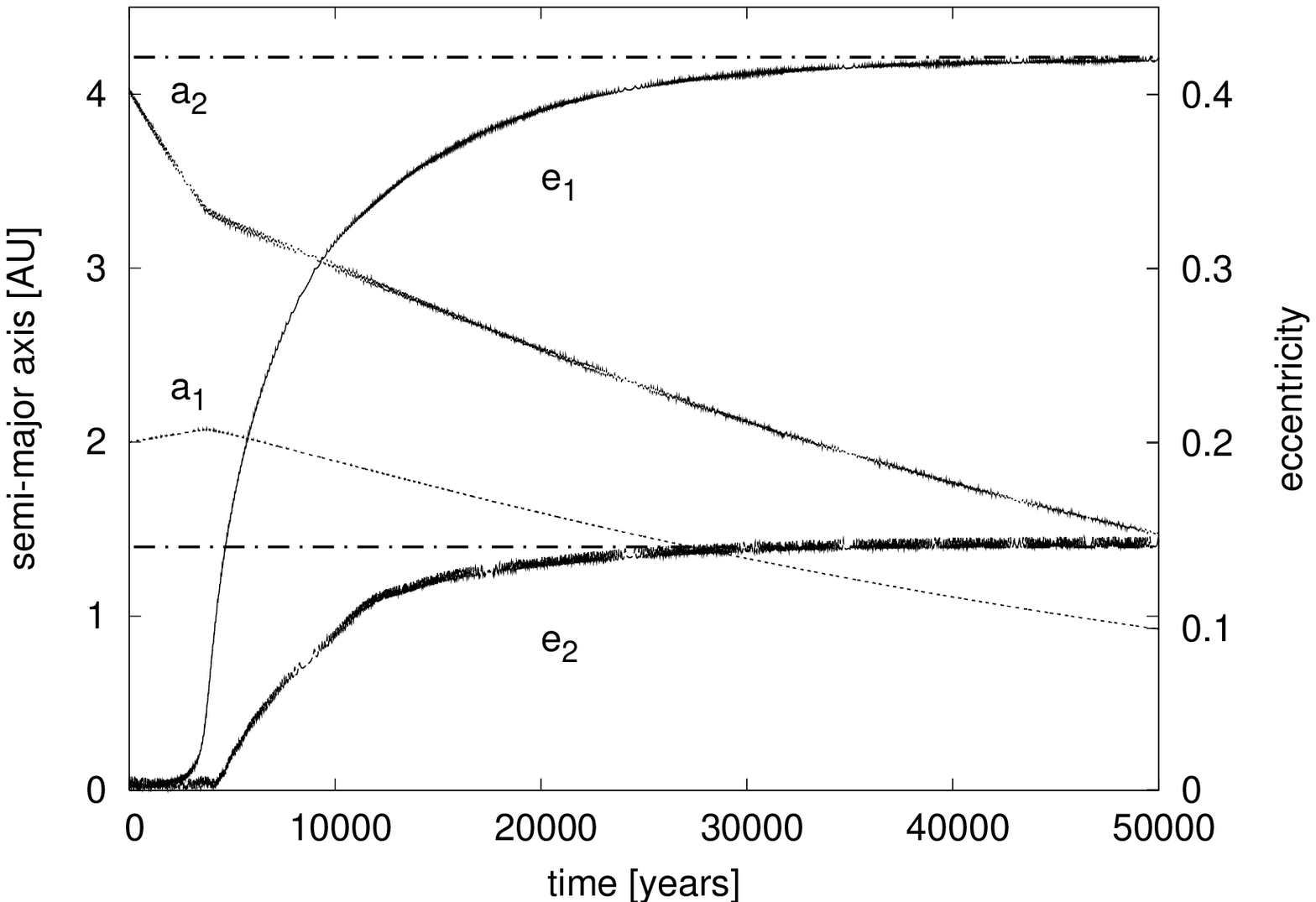}  
      \caption{Behaviour of the semi-major axes and the eccentricities
      of the giant planets in the resonant system HD~82\,943. The
      horizontal lines correspond to the observed values of the
      eccentricities. \emph{Top\,:} Only the motion of the outer
      planet is affected, with the $e-$folding times
      $\tau_{a_2}=2\times 10^4$ and $\tau_{e_2}=2.5\times 10^3$ years
      ($K_2=8$).  \emph{Bottom\,:} An inner disc is also assumed, with
      $\tau_{a_1}=-2\times 10^5$, $\tau_{e_1}=5\times 10^4$ years
      $e-$folding times for the inner planet ($K_1=4$), which allows
      the use of larger $\tau_{e_2}=2.5\times 10^3$ years and thus
      lower ratio $K_2 =4$ for the outer planet.  }
      \label{fig:hd829_diss_nbody}  
\end{figure}

\begin{figure}  
   \centering  
   \includegraphics[width=8cm]{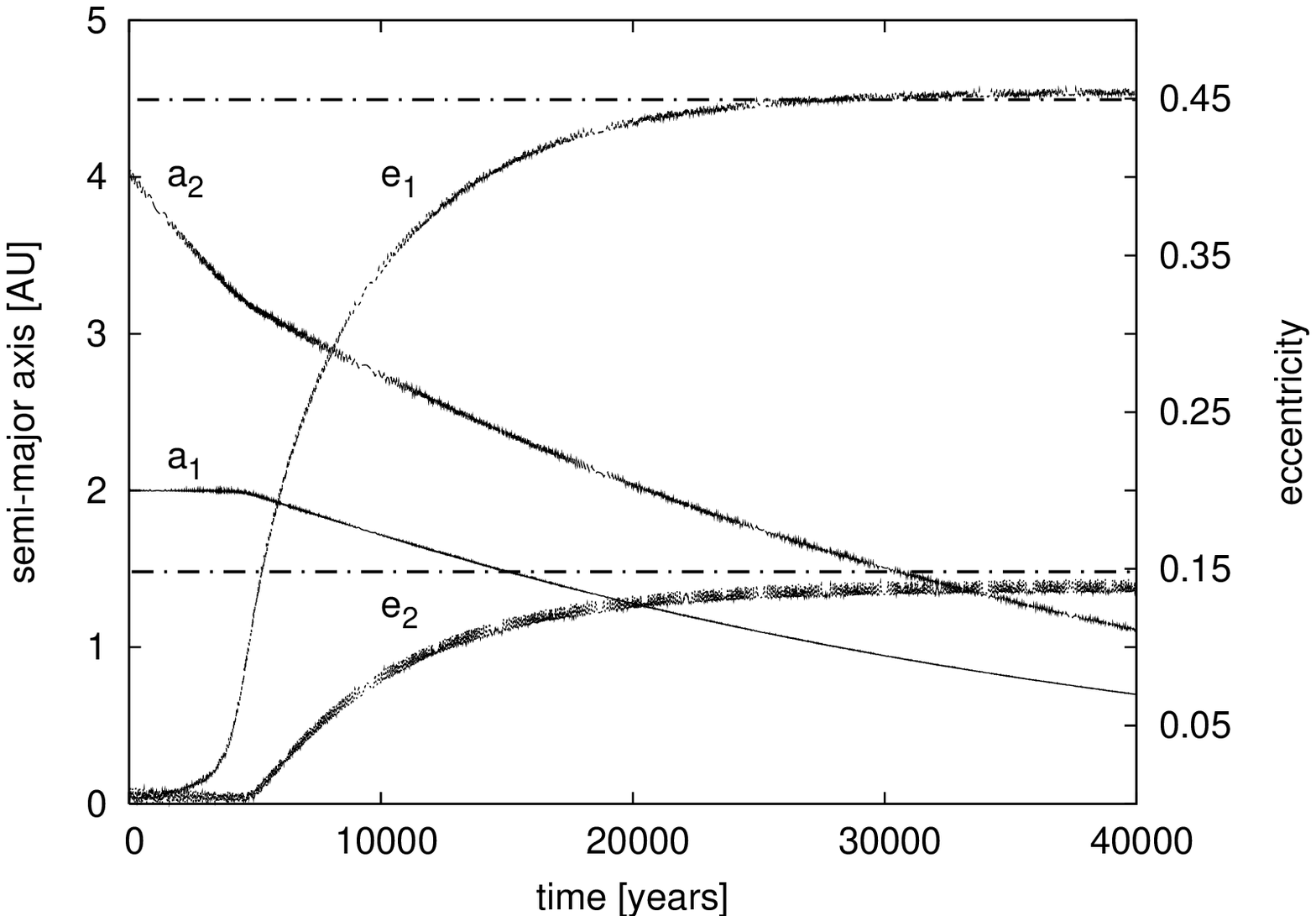}  
   \centering  
   \includegraphics[width=8cm]{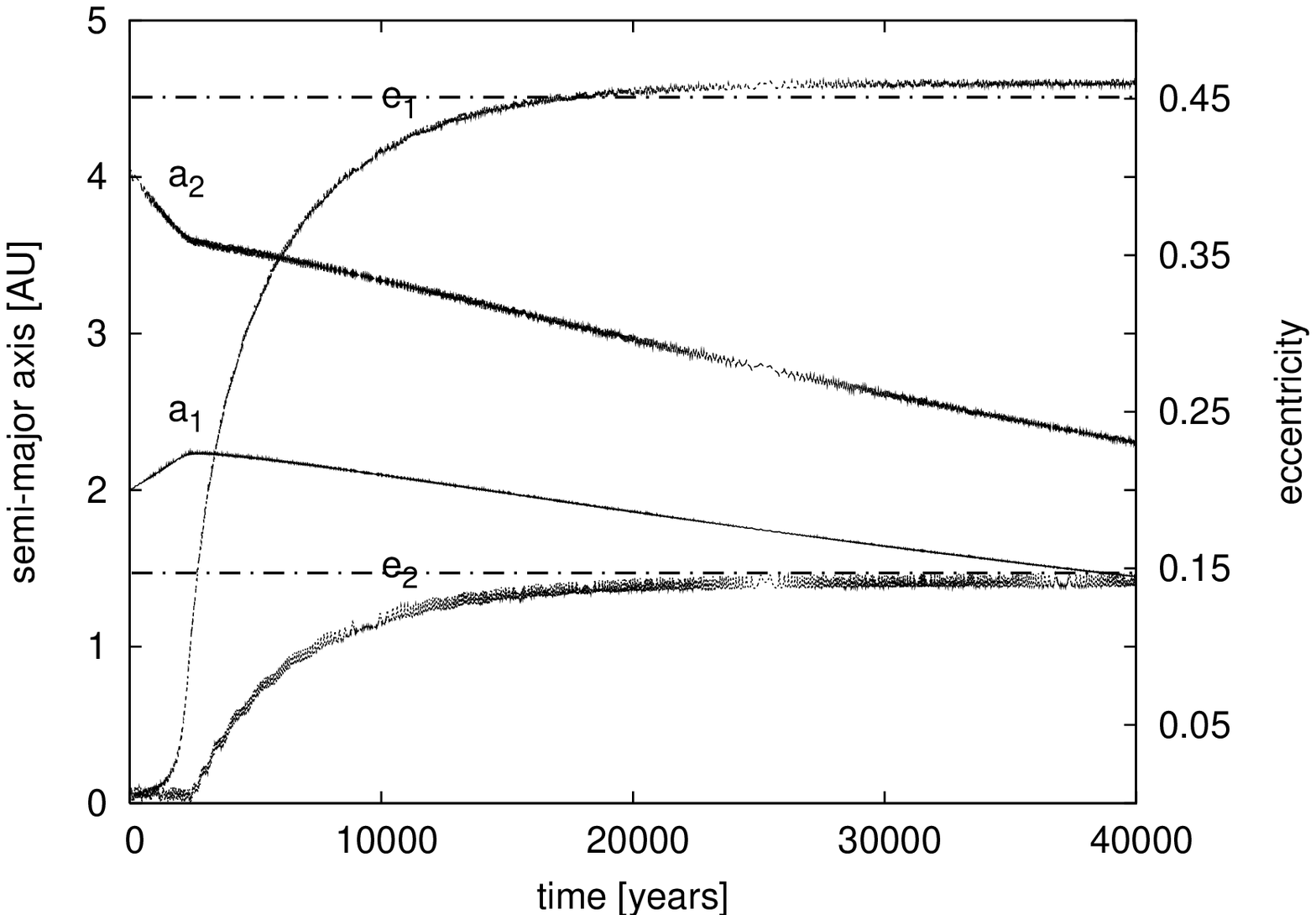}  
      \caption{ Behaviour of the semi-major axes and the
      eccentricities of the resonant giant planets in the resonant
      system HD~128\,311. The horizontal lines correspond to the
      observed values of the eccentricities. \emph{Top\,:} Only the
      motion of the outer planet is damped by an outer disc, with
      $K_2=\tau_{a_2}/\tau_{e_2}=7$.  \emph{Bottom\,:} An inner disc
      is also assumed. The $e-$folding times for the inner planet are
      $\tau_{a_1}=2\times 10^4$, $\tau_{e_1}=10^4$ years ($K_1=2$),
      which allows to use much lower ratio $K_2 =2$ for the outer
      planet.}
      \label{fig:hd128_diss_nbody}  
\end{figure}  

\paragraph{HD~82\,943\,:}
Recently \citet{2006ApJ...641.1178L} presented four sets of orbital
solutions based on a best-fit double-Keplerian model for
HD~82\,943. Using their orbital elements as initial conditions, direct
numerical integrations show that three of them exhibit ordered
behaviour, while one (Fit~I in the cited paper) is destabilised after
a few thousand years. In one of the three dynamically stable orbital
solutions (Fit~II) the variations of the eccentricities are
negligible, and the resonance variables oscillate with small
amplitudes indicating clearly that the system is deep in the 2:1
MMR. Since this behaviour can be reached during an inward convergent
migration of the giant planets, we investigate the formation of this
system using the scenario of the planetary migration.

First we assume that only an outer disc is present and the outer
planet feels its damping effect. To obtain the dynamical state
calculated based on Fit~II ($e_1=0.422$, $e_2=0.14$, $a_1=0.74$~AU,
$a_2=1.18$~AU), we needed the ratio $K_2 = 8$, using $\tau_{a_2}=2\times
10^4$ and $\tau_{e_2}=2.5\times 10^3$ years. This ratio $K_2$ does not
seem too high (as it lies typically between $1$ and $10$),
however, similarly to HD~73\,526, the eccentricities are slightly
increasing during the whole migration process. This is not a problem
if the migration of the planets is terminated gradually around the
observed values of their semi-major axes but it  may result in
exceeding the limits derived from observations, see the top panel of
Fig.~\ref{fig:hd829_diss_nbody}. On the other hand, by assuming the
presence of an inner disc, the eccentricities reach their constant
values gradually which do not seem to change further during the
migration process. The latter behaviour is shown in the bottom panel
of Fig.~\ref{fig:hd829_diss_nbody}. During this simulation we used for
the inner planet $\tau_{a_1}=-2\times 10^5$ and $\tau_{e_1}=5\times
10^4$ years, and for the outer planet $\tau_{a_2}=2\times 10^4$ and
$\tau_{e_2}=5\times 10^3$ years (giving $K_1=K_2=4$). Since the
eccentricities seem to reach constant values, this case seems to be a
more convenient formation scenario for HD~82\,943 than the previous
one.

We can conclude that the present dynamical behaviour of the resonant
system HD~82\,943 (based on Fit II of \citet{2006ApJ...641.1178L}) can
be explained in both ways\,: by assuming an outer disc alone or by
assuming an inner and an outer disc. However, the latter case seems
to be more reasonable since when the migration occurs over longer
times it guarantees constant eccentricity values during the whole
process.

\paragraph{HD~128\,311\,:}
Finally, we present our results for HD~128\,311. The most recent
orbital solution for this system is presented by
\cite{2005ApJ...632..638V}, while possible formation scenarios are
outlined by \cite{2006A&A...451L..31S}. In the latter study, the
formation of the resonant system is modelled by an inward convergent
migration of the giant planets which is followed by a sudden
perturbation\,; only the effect of an outer disc was considered.
Prior to the sudden perturbation, the outer giant planet migrated
inward and captured the inner one into a 2:1 resonance. It is
essential to find the final values of the eccentricities as due to the
migration process since the perturbative events modify only their
oscillations. These values are $e_1\approx 0.45$ and $e_2 \approx
0.15$ being obtained after a migration characterised by
$K_2=\tau_{a_2}/\tau_{e_2}=5$. In the present study we need somewhat
larger ratio, $K_2=7$, since contrary to \citet{2006A&A...451L..31S},
we do not stop the migration when the planets reach their actual
positions. The evolution of the semi-major axes and eccentricities are
presented in the top panel of Fig.~\ref{fig:hd128_diss_nbody}.

Assuming an inner disc, the same final state of the migratory evolution 
of HD~128\,311 can be obtained. In our numerical experiments we used 
$\tau_{a_1}=-2\times 10^4$, $\tau_{e_1}=10^4$ years for the inner planet, 
and $\tau_{a_2}=2\times 10^4$, $\tau_{e_2}=10^4$ years for the outer planet 
(being equivalent to $K_1=K_2=2$). The evolution of the system is shown in
the bottom panel of Fig.~\ref{fig:hd128_diss_nbody}. Comparing the
top and the bottom panels of Fig.~\ref{fig:hd128_diss_nbody}, we can
conclude that the formation of HD~128\,311 can also be modelled by
assuming an inner disc, however in this particular case, this is not
definitely necessary.

\subsection{Final state of a migration with an inner disc\,:
Numerical experiments}
\label{subsec:knum}

In this part we show the results of additional numerical simulations in
which we studied how the different parameters of migration influence
the final state of the resonant system. It was quite evident from the
beginning that the characterisation based only on the ratios $K_1$ and
$K_2$ would not yield unique results in the case when the inner planet
is also affected by an inner disc.

Our calculations are based on the observed orbital and physical
parameters of GJ~876, as given by \citet{2002ApJ...567..596L}. For the
inward migration of the outer planet we fixed $\tau_{a_2} = 4\times
10^4$, $\tau_{e_2} = 5\times 10^3$ years (so, $K_2=8$), and we changed
$\tau_{a_1}$ and $\tau_{e_1}$ in such a way to obtain the observed
state of the system.

\begin{figure}  
\centering  
\includegraphics[width=0.7\linewidth,angle=270]{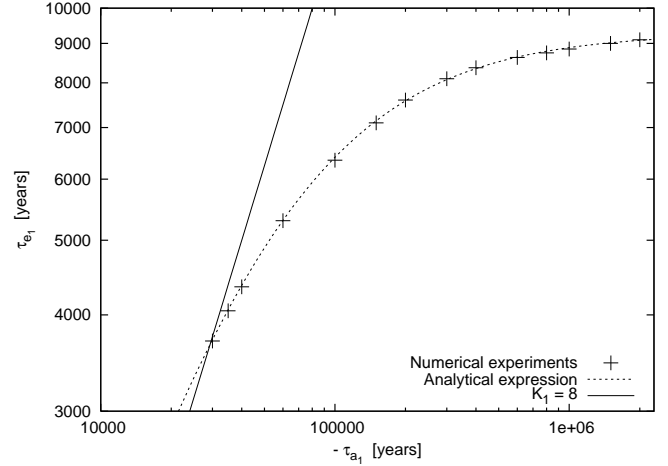}  
\caption{Pairs of migration parameters $\tau_{a_1}$ and $\tau_{e_1}$
         which result in the observed state of the system GJ~876 for
         fixed $\tau_{a_2}=4\times 10^4$ years and $\tau_{e_2}=5\times
         10^3$ years. \emph{Crosses\,:} Empirically found with N-body
         simulations. \emph{Dashed curve\,:} Analytical expression
         Eq.~(\ref{eq:tau_e_1}) (see
         Sect.~\ref{subsec:kanal}). \emph{Solid line\,:} Constant
         ratio $K_1=8$, for comparison.}
\label{fig:damping_parameters}  
\label{fig:damping_parameters_fit}  
\end{figure}

As starting values to our numerical simulations we used $\tau_{a_1} =
-3\times 10^4$ years and found that $\tau_{e_1} = 3.7\times 10^3$
years gives a correct result ($K_1\approx 8$ here). Then we increased
gradually the absolute value of $\tau_{a_1}$ corresponding to a weaker
damping on $a_1$. To obtain the observed behaviour of the system, for
larger $\tau_{a_1}$, we needed larger $\tau_{e_1}$ meaning a weaker
damping on $e_1$ too. In order to explore the mutual dependence of
$\tau_{a_1}$ and $\tau_{e_1}$, we performed a series of numerical
experiments. Our results are shown in
Fig.~\ref{fig:damping_parameters}\,; on the $x-$axis the absolute
value of $\tau_{a_1}$, while on the $y-$axis the eccentricity damping
time $\tau_{e_1}$ are displayed in logarithmic scale. The crosses show
those corresponding values of $\tau_{a_1}$ and ${\tau_{e_1}}$ which
are needed to obtain the same final values of the eccentricities $e_1$
and $e_2$ ($0.26$ and $0.035$ respectively). The $\tau_{e_1}$
increases rapidly for small $|\tau_{a_1}|$, however it is not
proportional to $\tau_{a_1}$ (compare with the straight line on
Fig.~\ref{fig:damping_parameters}), but it levels off and clearly tends
to a limit $\tau_{e_1 \rm max}$, which in this case is slightly higher
than $9\times 10^3$ years.

We can conclude that for a fixed pair of $\tau_{a_2}$ and
$\tau_{e_2}$, there is no unique $K_1$ that determines the final state
of the system. On the contrary, there exists a $\tau_{e_1 \rm max}$,
which determines the final state of the system if the inner disc does
not affect the semi-major axis of the inner planet. In reality,
however, the inner disc (as rotating faster) can transmit angular
momentum to the inner planet as well as energy, increasing its
semi-major axis. In this case we need smaller $\tau_{e_1}$, or
equivalently, more effective damping on $e_1$.

If we compare the two migration scenarios\,: (i) only an outer disc is
present and there is no inner disc between the inner planet and the
star\,; and (ii) both an outer disc and an inner disc are present and
the planets orbit in a common gap between them, we can summarise the
following results\,: The final state of the system in case (i) can be
described solely by the ratio $K_2=\tau_{a_2}/\tau_{e_2}$. In case
(ii), the ratios $K_1$ and $K_2$ are not enough, the final
state of the system depends on additional parameters too\,; in the
most general case, when the inner disc forces the inner planet to
migrate outward, there are essentially three parameters, which can be
$K_1$, $K_2$, and the ratio $\tau_{a_1}/\tau_{a_2}$ for instance. In
next subsection, we analyse theoretically this behaviour.

\subsection{Final state of a migration with an inner disc\,: Analytics}
\label{subsec:kanal}

For the semi major axis and eccentricity of a planet to
decrease exponentially with damping timescales $\tau_a=-\dot{a}/a$ and
$\tau_e=-\dot{e}/e$, its energy $E$ and angular momentum $H$ must vary with
the following rates (through Eqs.~(\ref{eq:adot}) and
(\ref{eq:edot})~)\,:
\begin{eqnarray}
\label{eq:Edot}
\dot{E} & = & E/\tau_a\\
\dot{H} & = & \frac{H}{2}
\left(\frac{2e^2}{1-e^2}\frac{1}{\tau_e}-\frac{1}{\tau_a}\right)
\label{eq:Hdot}
\end{eqnarray}

When two planets are considered, the variation of the energy of the
system \{~pair of planets~\}, $\dot{E}$ is the sum of the energy
variations applied to both planets, so that
\begin{equation}
\dot{E} = \frac{E_1}{\tau_{a_1}}+\frac{E_2}{\tau_{a_2}}\ .
\label{eq:dotEtot}
\end{equation}
If the two planets are in resonance, the ratio between their semi
major axes is kept constant, and consequently also the ratio
between their energies. Let us define\,:
\begin{equation}
\varepsilon = E_2/E_1 = M_2a_1/M_1a_2\ .
\label{eq:eps}
\end{equation}
Then,
\begin{equation}
E = E_1+E_2 = (1+\varepsilon)E_1\ ,
\label{eq:E1}
\end{equation}
\begin{equation}
\dot{E} = (1+\varepsilon) \dot{E_1}\ .
\label{eq:dotE1}
\end{equation}
Note that due to some energy exchange between the planets through the
resonance, the variation of the energy of the first planet is not
$E_1/\tau_{a_1}$, but $\dot{E_1}=\dot{E}/(1+\varepsilon)$, with
$\dot{E}$ given by Eq.~(\ref{eq:dotEtot}).

\ 

The same holds for the angular momentum. For the angular
momentum $H$ of the system \{~pair of planets~\} ($H=H_1+H_2$), the total
variation rate reads (from Eq.~(\ref{eq:Hdot})~)\,:
\begin{equation}
\dot{H} = \sum_{i=1,2}\frac{H_i}{2}
\left(\frac{2{e_i}^2}{1-{e_i}^2}\frac{1}{\tau_{e_i}}-\frac{1}{\tau_{a_i}}\right)\ .
\label{eq:dotHtot}
\end{equation}
For the final state, when the planets are in resonance and their
eccentricities are also constant, as is the ratio of their angular
momentum. Let us define\,:
\begin{equation}
\eta = H_2/H_1 = \frac{M_2}{M_1}\sqrt{\frac{a_2(1-{e_2}^2)}{a_1(1-{e_1}^2)}}\ .
\label{eq:eta}
\end{equation}
Then, $H=H_1+H_2=(1+\eta)H_1$, and also
\begin{equation}
\dot{H} = (1+\eta)\dot{H_1}\ .
\label{eq:dotH1}
\end{equation}

If the eccentricities are constant ($\dot{e_i}=0$), then, from
Eqs.~(\ref{eq:edot}), (\ref{eq:E1}), (\ref{eq:dotE1}), and finally
(\ref{eq:dotEtot})\,:
\begin{equation}
\dot{H_i} = -\frac{H_i}{2}\frac{\dot{E_i}}{E_i} =
-\frac{H_i}{2}\frac{\dot{E}}{E} =
-\frac{H_i}{2}\left(\frac{1/\tau_{a_1}+\varepsilon/\tau_{a_2}}{1+\varepsilon}\right)
\label{eq:dotHi}
\end{equation}

Using Eqs.~(\ref{eq:dotHtot}) and (\ref{eq:dotHi}),
Eq.~(\ref{eq:dotH1}) gives a relation between $\tau_{a_1}$,
$\tau_{a_2}$, $\tau_{e_1}$, and $\tau_{e_2}$. In the problem that we
study, the masses of the planets are known, as well as their
eccentricities and the ratio between their semi major axes. The
question is\,: for a given effect on the outer planet, what should the
effect on the inner planet be to be consistent with the observed
eccentricities\,? Below, we solve this equation in the unknown
$\tau_{e_1}$ with the free parameter $\tau_{a_1}$, while $\tau_{a_2}$,
$\tau_{e_2}$, $\varepsilon$ and $\eta$, are given constants
($\varepsilon$ and $\eta$ being defined by Eqs.~(\ref{eq:eps}) and
(\ref{eq:eta}) respectively). Using Eqs.~(\ref{eq:dotHtot}) and
(\ref{eq:dotHi}), Eq.~(\ref{eq:dotH1}) reads\,:
$$\sum_{i=1,2}\frac{H_i}{2}
\left(\frac{2{e_i}^2}{1\!-\!{e_i}^2}\frac{1}{\tau_{e_i}}-\frac{1}{\tau_{a_i}}\right) =  -(1+\eta)\frac{H_1}{2} \left(\frac{1/\tau_{a_1}+\varepsilon/\tau_{a_2}}{1+\varepsilon}\right)$$
$$\frac{2{e_1}^2}{1\!-\!{e_1}^2}\frac{1}{\tau_{e_1}}
- \frac{1}{\tau_{a_1}}
+ \eta\left(\frac{2{e_2}^2}{1\!-\!{e_2}^2}\frac{1}{\tau_{e_2}}
- \frac{1}{\tau_{a_2}}\right)
 =
- \frac{1\!+\!\eta}{1\!+\!\varepsilon} \left(\frac{\varepsilon}{\tau_{a_2}}+\frac{1}{\tau_{a_1}}\right)$$
$$\frac{2{e_1}^2}{1\!-\!{e_1}^2}\frac{1}{\tau_{e_1}} = \frac{1}{\tau_{a_2}}\left(\eta-\frac{(1\!+\!\eta)\varepsilon}{1+\varepsilon}\right) - \frac{1}{\tau_{e_2}}\frac{2{e_2}^2\eta}{1-{e_2}^2} + \frac{1}{\tau_{a_1}}\left(1-\frac{1\!+\!\eta}{1\!+\!\varepsilon}\right)\ .$$
Eventually, we obtain\,:
\begin{equation}
\frac{1}{\tau_{e_1}} = \underbrace{\frac{1\!-\!{e_1}^2}{2{e_1}^2} \left( \frac{1}{\tau_{a_2}}\frac{\eta-\varepsilon}{1+\varepsilon} - \frac{1}{\tau_{e_2}}\frac{2{e_2}^2\eta}{1-{e_2}^2}\right) }_{1/\tau_{e_1 \rm max}} + \frac{1}{\tau_{a_1}}\underbrace{\frac{1\!-\!{e_1}^2}{2{e_1}^2}\frac{\varepsilon-\eta}{1+\varepsilon}}_{K_{1,0}}
\label{eq:tau_e_1}
\end{equation}

As one can see, the damping rate that should be applied to the inner
planet ($1/\tau_{e_1}$) is the sum of two terms.

The first one ($1/\tau_{e_1 \rm max}$) is required to balance the
action on the outer planet\,; indeed, when no force is applied to the
inner planet, the energy loss rate of the outer planet is not the
expected $E_2/\tau_{a_2}$, but $(E_2/\tau_{a_2})
\frac{\varepsilon}{1+\varepsilon}$ because the inner planet also has
to lose energy to preserve the resonant motion\,; thus, the angular
momentum loss rate is overestimated by the expression
Eq.~(\ref{eq:Hdot}), and both eccentricities rise. Therefore, a
damping of the eccentricity needs to be applied on the inner planet
as well.

The second term ($K_{1,0}/\tau_{a_1}$) is proportional to
${\tau_{a_1}}^{\!-1}$\,; the coefficient $K_{1,0}$ is negative if
${e_1}^2>(1-(a_2/a_1)^3)+(a_2/a_1)^3{e_2}^2$, which is always true in
the case of a 2:1 MMR for $e_2<0.866$.


In the case studied in the previous Sect.~\ref{subsec:knum}, one
has \citep{2002ApJ...567..596L}\,: $M_2=1.808 M_{\rm Jup}=5.65\times
10^{-3}M_*$ , $M_1=0.5696 M_{\rm Jup}=1.78\times 10^{-3}M_*$ ,
$e_1\approx 0.265$ , $e_2\approx 0.035$ , $\tau_{a_2}=4\times 10^4$ ,
$\tau_{e_2}=5\times 10^3$ and the two planets are in 2:1 MMR, with
$a_2/a_1=1.602$. Thus, $\eta=4.1641$ and $\varepsilon=1.9812$. This
gives $\tau_{e_1 \rm max}=9289$ years, and $K_{1,0}=-4.8471$. The
dashed curve on Fig.~\ref{fig:damping_parameters_fit} shows
$\tau_{e_1}$ as a function of $\tau_{a_1}$ from
Eq.~(\ref{eq:tau_e_1}). The fit is excellent.

This shows that Eq.~(\ref{eq:tau_e_1}) efficiently gives what
parameters one should choose to reproduce a given system by the means
of N-body simulations with dissipative forces, or one should have in
the protoplanetary disc. In particular, the expression of $\tau_{e_1
\rm max}$ shows that if $e_2$ is relatively small, a huge $K_2$ of the
order of ${e_2}^{-2}$ is required for $\tau_{e_1 \rm max}\to
\infty$. If the inner disc tends to make the inner planet migrate
outward with $\tau_{a_1}<0$, an additional damping on its eccentricity
is required, which can be expressed as $\tau_{e_1 \rm
add}=\tau_{a_1}/K_{1,0}$, with $K_{1,0}<0$, depending on all the
parameters of the system. On the other hand, if the inner disc tends
to attract the planet sufficiently (which may happen as has been shown
in Sect.~\ref{sec:ID}), no eccentricity damping on the inner planet
may be required at all (this happens in our case if $\tau_{a_1} =
+45025$ years).

With this equation, one can try to find reasonable parameters (with
not too big $K_1$ and $K_2$) to explain how all the known resonant
exoplanetary systems can have been formed in the disc. Our numerical
simulations have shown that this can be achieved for at least 4 such
systems.

\section{Conclusion}
\label{sec:conclusion}

In this paper, we address the problem of the moderate eccentricity of
extra-solar planets in resonance. The resonant configuration requires
a convergent migration of the planets, but continued migration in
resonance leads to unlimited eccentricity growth if no eccentricity
damping mechanism is at work.

In Sect.~\ref{sec:ID}, we have shown that an inner disc has a non
negligible influence on a giant planet on an eccentric orbit and
modifies its orbital parameter.  The strength of this effect varies
with planetary eccentricity.  For very small eccentricities, $e \leq
0.05$, we find a very small but positive $\dot{e}$ while for larger
$e$ it becomes more and more negative.  The induced change in
semi-major axis remains positive, as expected for the Lindblad torques
induced by an inner disc on a massive planet.  Only for larger
eccentricities $e\geq 0.35$ the change in semi-major axis becomes
negative, leading to inward migration.  The usage of a constant
$K$-factor between the $e$-folding time scales of the semi-major axis
and the eccentricity is clearly an over simplification.

However, the measured influence of the disc on the planet depends on
the adopted tapering function applied to exclude at least parts of the
gas within the Hill sphere of the planet, and possibly on other
numerical effects, such as the smoothing length, the resolution, or
boundary conditions. Hence, determination of the exact, absolute
magnitude of the effect is a bit difficult.

Anyway, an eccentricity damping should be applied to the inner planet
to obtain realistic results, in particular if its eccentricity is
larger than $0.1$.  Using a hybrid 2D-1D hydro-code that allows the
simulation of the whole disc from its physical inner boundary to an
arbitrary large radius, we compute the longterm evolution of the GJ~876
system, for which the orbital elements are well known, and the radius
of the inner edge of the disc can be estimated thanks to the presence
of a third planet very close to the star. We find that the inner disc
does not disappear after the planets opened a common gap, and that it
effectively helps damping the eccentricity of the inner planet. With
realistic disc parameters (viscosity and aspect ratio), we are able to
reproduce the observations.

As hydro-simulations are very time-consuming, we finally perform
customised N-body simulations with explicit non-conservative forces,
added to mimic the effect of the outer and inner disc. We find that
applying a significant damping to both planets, as required by the
influence of both the inner and outer discs, enables us to fit quite
well a few other exoplanetary systems. In addition, N-body simulations
have shown that when two planets are considered, the ratio between the
eccentricity and semi major axis damping applied on each of them
($K_1$ and $K_2$) are not sufficient to determine the final state of
the system. In fact, for a given damping in $a$ and $e$ applied on the
outer planet, one can express analytically the eccentricity damping
that should be applied on the inner planet to match the observed
orbital configuration, as a function of its semi major axis damping.

Given the satisfying fits of a few systems that we obtain, we claim
that the problem of the low eccentricities of the resonant
exoplanetary systems simply stems from the fact that the inner disc
had not been taken properly into account, which is not reasonable. A
pair of planets orbiting in a protoplanetary disc may well orbit in a
common gap and enter a Mean Motion Resonance, but it does not
necessarily open a complete cavity from the star to the outermost
planet, so that the inner disc should influence the planets
dynamics. \citet{Crida-Morby-2007} already suggested that the opening
of such a cavity by a giant planet requires very specific
conditions\,; in fact, according to our hydro and N-body simulations,
the low observed eccentricities of the exoplanets in resonance support
the idea that the inner disc does not disappear in general.

\begin{acknowledgements}

A. Crida acknowledges the support through the German Research
Foundation (DFG) grant KL 650/7.  Zs. S\'andor thanks the supports of the
Hungarian Scientific Research Fund (OTKA) under the grant PD48424 and
of the DFG under the grant 436 UNG 17/1/07.

Very fruitful discussions with F. Masset and C. Dullemond are
gratefully acknowledged.

\end{acknowledgements}

\bibliographystyle{aa}
\bibliography{crida,kley8}
\end{document}